%% file: main.tex
\documentclass[11pt]{article}

\usepackage[preprint]{acl}
\usepackage[T1]{fontenc}
\usepackage[utf8]{inputenc}
\usepackage{times}

\usepackage{latexsym}
\usepackage{microtype}
\usepackage{graphicx}
\usepackage{amsmath,amssymb}
\usepackage{booktabs}
\usepackage{multirow}
\usepackage{array}
\usepackage{colortbl}
\usepackage{pifont}
\usepackage{makecell}
\usepackage{stfloats}
\usepackage{float}
\usepackage{afterpage}
\usepackage{tcolorbox}
\tcbuselibrary{breakable}
\newcolumntype{C}{c}
\newcommand{\PaperLanguageLabelPrefix}{en:}
\title{SWE-Time: Benchmarking Temporal Repository Understanding}
\author{\normalfont\normalsize
\begin{tabular}{@{}c@{}}
{\bfseries Shijun Xiao$^{1,3,\dagger}$, Xinrong Chen$^{2,\dagger}$, Hao Liu$^{3}$, Wensheng Zhang$^{4}$, Yongqiang Tang$^{3,*}$}\\[6pt]
$^1$Nankai University, Tianjin, China\qquad $^2$Baidu, Inc., Beijing, China\\[2pt]
$^3$Institute of Automation, Chinese Academy of Sciences, Beijing, China\\[2pt]
$^4$Guangzhou University, Guangzhou, China\\[2pt]
$^\dagger$Equal contribution.\\[2pt]
$^*$Corresponding author: \texttt{yongqiang.tang@ia.ac.cn}
\end{tabular}}
\hypersetup{pdftitle={SWE-Time: Benchmarking Temporal Repository Understanding},pdfauthor={Shijun Xiao, Xinrong Chen, Hao Liu, Wensheng Zhang, Yongqiang Tang}}

\begin{document}
\maketitle

\input{sections/english/abstract}
\input{sections/english/introduction}
\input{sections/english/related_work}

\input{sections/english/method}

\input{sections/english/experiments}
\input{sections/english/conclusion}
\input{sections/english/limitations}

\bibliography{custom}
\clearpage
\appendix
\renewcommand{\theHsection}{appendix.\Alph{section}}
\renewcommand{\theHsubsection}{appendix.\Alph{section}.\arabic{subsection}}
\renewcommand{\theHsubsubsection}{appendix.\Alph{section}.\arabic{subsection}.\arabic{subsubsection}}
\input{sections/appendix_case_cards}

\input{sections/english/appendix_a}
\input{sections/english/appendix_b}
\input{sections/english/appendix_c}
\input{sections/english/appendix_d}
\end{document}

%% file: sections/english/abstract.tex
\begin{abstract}
Repository-level code-understanding benchmarks have progressively expanded the spatial scope of reasoning, from local code to entire repositories, while largely treating each repository as a static snapshot. However, how models reason about repository evolution over time remains largely unexplored. We introduce SWE-Time, a benchmark for temporal repository understanding that makes repository evolution itself the reasoning target. SWE-Time evaluates two complementary temporal directions: retrospective understanding of repository evolution through historical regression analysis and prospective understanding of intended changes from pre-change repository states. To enable realistic yet reliable evaluation in these settings, we introduce the RealWorld-to-Verifiable Construction Framework, which transforms real-world developer queries and historical evidence into verifiable evaluation tasks. We evaluate 15 coding agents on 1,661 cases from 65 GitHub repositories. Within Claude Code, Claude Opus 4.8 achieves the highest SWE-Time Avg. of 61.9, leaving substantial headroom. Controlled analyses highlight the importance of historical code evidence and developer intent, while consistent reasoning across repository states remains challenging. These results highlight repository evolution as a distinct and challenging dimension of code understanding.
\end{abstract}

%% file: sections/english/introduction.tex
\begingroup\input{figures/figure_1}\endgroup

\section{Introduction}
\label{en:sec:introduction}

\begin{flushright}
\small
\textit{``Why do we remember the past but not the future?''}\\[0.3em]
---Stephen Hawking, \textit{A Brief History of Time}
\end{flushright}

\noindent Coding has become one of the most important applications of large language models, motivating numerous benchmarks for code generation~\citep{chen2021humaneval,apps2021apps,liu2023evalplus,jain2025livecodebench,zhuo2025bigcodebench}, software repair~\citep{jimenez2024swebench}, and repository-level reasoning~\citep{wu2024repobench,ding2023crosscodeeval,li2024deveval}. Despite their different objectives, these tasks share a common requirement: code understanding (CU), the ability to reason about program behavior and dependencies~\citep{yang2024sweagent,li2025longcodeu,wang2026reporeasoner,lu2021codexglue,wang2024codescope,nguyen2025codemmlu}. Dedicated CU benchmarks assess this shared capability directly at the repository level. For example, SWE-QA-Pro and RepoProbe use natural-language questions about repository code~\citep{cai2026sweqapro,yang2026repoprobe}, whereas RepoReason and RepoReasoner evaluate assertion verification, output prediction, and call-chain prediction~\citep{li2026reporeason,wang2026reporeasoner}.

These benchmarks expand the spatial scope of CU evaluation from individual functions and files to entire repositories. However, they generally assess code within a fixed repository snapshot. In practice, each snapshot is part of an evolving repository history, as shown in Figure~\ref{en:fig:repository_evolution}. Current behavior reflects past changes, while the effects of a new change depend on the current repository state~\citep{tip2004chianti}. For example, when a feature that previously worked starts failing, developers need to identify which change introduced the regression~\citep{zeller1999yesterday}. Before implementing a request, developers need to anticipate its effects on repository components and behavior, even though the modified code does not yet exist~\citep{sillito2006questions,ko2007informationneeds}. These problems motivate evaluating both \textit{retrospective reasoning} about the origins of current behavior and \textit{prospective reasoning} about the consequences of future changes.

To assess these two forms of reasoning, we introduce \textit{SWE-Time}, a benchmark for \textit{temporal repository understanding}. Its two tasks instantiate these reasoning directions. In \textit{Task Backward}, agents trace a reported regression through repository history to identify the first-bad commit. In \textit{Task Forward}, they reason from developer intent and the pre-change repository state to predict how the requested change would affect repository behavior. They assess whether agents can relate code changes to repository behavior in both directions.

Realistic, reproducible evaluation requires authentic developer queries and verifiable reference answers~\citep{just2014defects4j,zhang2020bugsinpy,gyimesi2019bugsjs,tomassi2019bugswarm}. Existing CU benchmarks fall into two categories: natural-language QA benchmarks and programmatic benchmarks. QA benchmarks~\citep{cai2026sweqapro,yang2026repoprobe} offer developer-oriented questions but often score open-ended answers with LLM judges~\citep{zheng2023judgingllm}. Programmatic benchmarks~\citep{li2026reporeason,wang2026reporeasoner}, by contrast, support deterministic scoring but typically do not start from natural-language developer requests. We introduce the RealWorld-to-Verifiable Construction Framework, using independently verifiable historical evidence to identify authentic development events suitable for reliable evaluation. The resulting tasks combine authentic developer queries with deterministic answer verification.

Applying this procedure yields 1,661 cases from 65 GitHub repositories. We evaluate 15 coding agents on the resulting benchmark, using Claude Code for the main model comparison and Codex for supplementary references. Claude Opus 4.8 achieves the highest SWE-Time Avg. within Claude Code at 61.9. To examine the sources of difficulty, we conduct controlled experiments on historical code evidence and developer intent, and assess correctness across paired repository states using \textit{Pair Exact}. These analyses show that both historical evidence and developer intent contribute to performance, while maintaining correct predictions across repository states remains difficult. Our contributions are as follows:

\begin{list}{\labelitemi}{}
\item We formulate temporal repository understanding as retrospective and prospective reasoning, and introduce SWE-Time to evaluate both through repository-grounded tasks.
\item We introduce the RealWorld-to-Verifiable Construction Framework, using independently verifiable historical evidence to identify authentic development events suitable for reliable evaluation.
\item We evaluate 15 coding agents and use controlled experiments to characterize the roles of historical evidence, developer intent, and repository state in their performance.
\end{list}

%% file: figures/figure_1.tex
\begin{figure}[!t]
\centering
\includegraphics[width=\columnwidth]{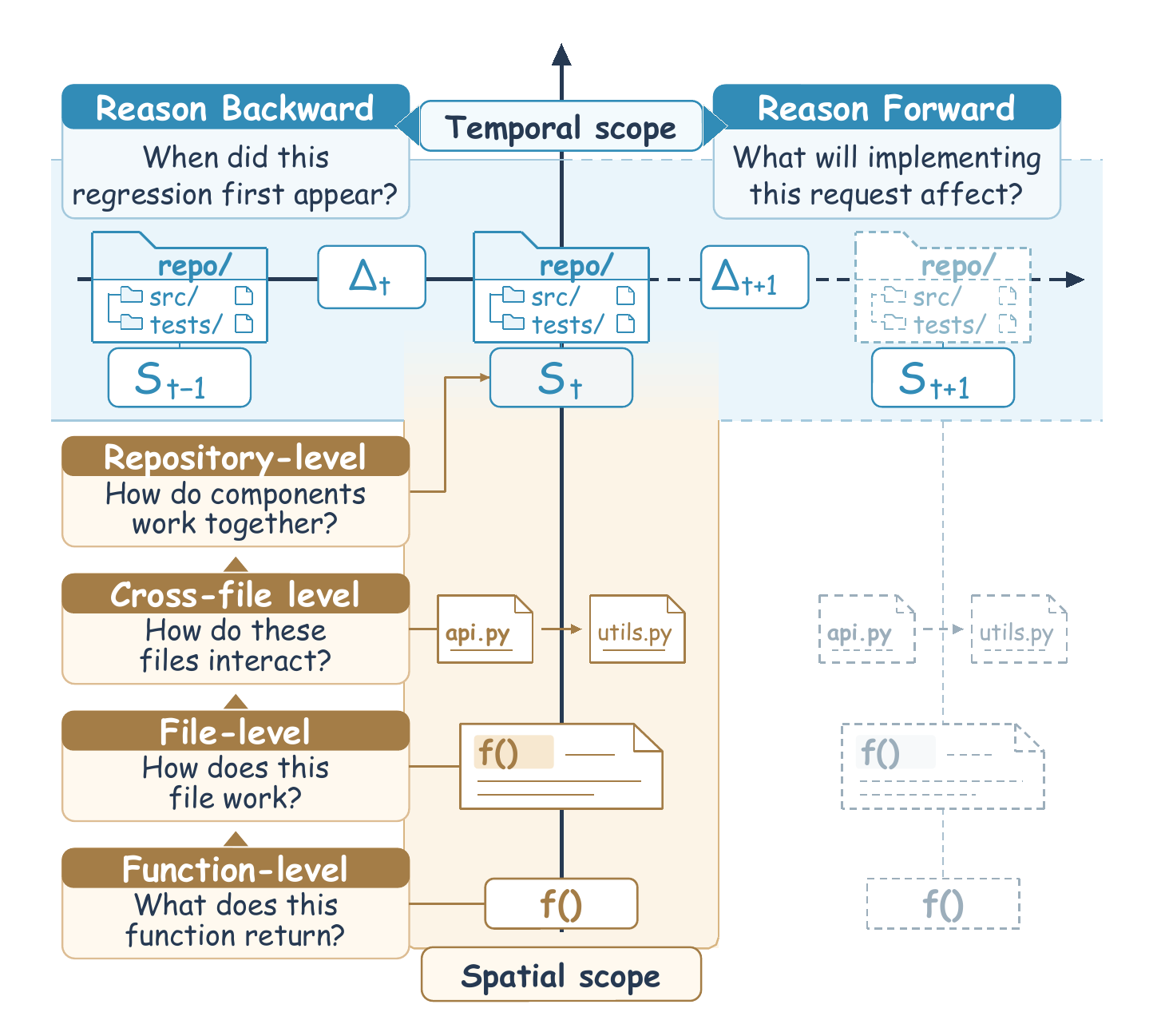}
\captionsetup{labelfont=bf}
\caption{Existing code-understanding benchmarks focus on a fixed repository snapshot, whereas SWE-Time evaluates how code evolves through retrospective and prospective reasoning. Spatial scope increases from functions to files, cross-file interactions, and repositories. $S_{t-1}$, $S_t$, and $S_{t+1}$ denote repository states, connected by changes $\Delta_t$ and $\Delta_{t+1}$. Dashed future states are hidden from the agent.}
\label{\PaperLanguageLabelPrefix fig:repository_evolution}
\end{figure}

%% file: sections/english/related_work.tex
\section{Related Work}
\label{en:sec:related_work}

\begingroup\input{tables/table_1}\endgroup

\textbf{Spatial Repository Understanding.}
CRUXEval evaluates function-level reasoning through input and output prediction~\citep{gu2024cruxeval}, while LongCodeU covers structure, data flow, and dependencies in long code contexts~\citep{li2025longcodeu}. At the repository level, SWE-QA and CoReQA transform information from development activities into questions~\citep{peng2026sweqa,chen2025coreqa}; SWE-QA-Pro synthesizes questions from issues and code~\citep{cai2026sweqapro}; and the CU track of SWE-Compass generates code-analysis tasks~\citep{xu2025swecompass}. RepoProbe draws on GitHub Discussions, with some queries rewritten by an LLM, and uses an LLM judge to score fact-level checklists~\citep{yang2026repoprobe}. These tasks offer developer-oriented inputs, but query provenance and scoring determinism differ across benchmarks.

Programmatic benchmarks instead use fixed targets: RepoReason evaluates assertion verification~\citep{li2026reporeason}, and RepoReasoner evaluates output and call-chain prediction~\citep{wang2026reporeasoner}. Their scoring does not require an LLM judge, but their inputs are not authentic natural-language developer requests. These code-understanding tasks primarily assess a given code context. SWE-Time additionally makes changes between repository states part of the evaluation target.

\textbf{Temporal Software Engineering.}
SZZ traces bug-fixing changes to historical changes that may have introduced the problem~\citep{sliwerski2005szz,williams2008szzrevisited,rosa2021szzoracle}. LLMBisect and AgentSZZ use LLMs or agents to reason over fixing evidence, code, and commit history to identify bug-inducing commits~\citep{zhang2026llmbisect,lyu2026agentszz}. LinkAnchor recovers resolving-commit links from issues and repository history~\citep{akhavan2026linkanchor}. SWE-EVO uses original release notes and linked developer artifacts to specify long-horizon software evolution tasks, validating resulting code changes with tests~\citep{le2026sweevo}. These studies already use software history and involve code and temporal reasoning. SWE-Time differs in its evaluation setting: Backward starts from a sanitized regression report and locates the first-bad boundary without future fixing evidence, while Forward infers workflow consequences from a pre-change issue and repository state. Its outputs directly assess historical localization and change consequences without requiring a repair patch.
Table~\ref{en:tab:benchmark_comparison} summarizes task, input, scoring, and temporal properties.

%% file: tables/table_1.tex
\providecommand{\cmark}{\ding{51}}   %
\providecommand{\xmark}{\ding{55}}   %
\providecommand{\pmark}{$\triangle$} %

\begin{table*}[t]
\centering
\setlength{\tabcolsep}{4.0pt}
\renewcommand{\arraystretch}{0.95}

\resizebox{\textwidth}{!}{
\begin{tabular}{@{}llccccc@{}}
\toprule

\makecell[c]{\textbf{Benchmark}}
&
\makecell[c]{\textbf{Primary Task}}
&
\makecell[c]{\textbf{Repo-Level?}}
&
\makecell[c]{\textbf{Real Developer}\\\textbf{Query?}}
&
\makecell[c]{\textbf{Verifiable?}}
&
\makecell[c]{\textbf{Code}\\\textbf{Understanding?}}
&
\makecell[c]{\textbf{Temporal?}}
\\

\midrule

CRUXEval
& Function-Level Code Reasoning
& \xmark & \xmark & \cmark & \cmark & \xmark \\

LongCodeU
& Long-Code Understanding
& \xmark & \xmark & \cmark & \cmark & \xmark \\

SWE-QA
& Repository QA
& \cmark & \pmark & \xmark & \cmark & \xmark \\

CoReQA
& Repository QA
& \cmark & \pmark & \xmark & \cmark & \xmark \\

SWE-QA-Pro
& Repository QA
& \cmark & \pmark & \xmark & \cmark & \xmark \\

SWE-Compass CU
& Code Understanding
& \cmark & \pmark & \xmark & \cmark & \xmark \\

RepoProbe
& Repository QA
& \cmark & \pmark & \xmark & \cmark & \xmark \\

RepoReason
& Assertion Verification
& \cmark & \xmark & \cmark & \cmark & \xmark \\

RepoReasoner
& Output / Call-Chain Prediction
& \cmark & \xmark & \cmark & \cmark & \xmark \\

LLMBisect
& Bug-Introducing Commit Localization
& \cmark & \xmark & \cmark & \xmark & \cmark \\

AgentSZZ
& Bug-Introducing Commit Localization
& \cmark & \xmark & \cmark & \xmark & \cmark \\

LinkAnchor
& Issue--Commit Linking
& \cmark & \cmark & \cmark & \xmark & \cmark \\

SWE-EVO
& Long-Horizon Software Evolution
& \cmark & \cmark & \cmark & \xmark & \cmark \\

\midrule

\textbf{SWE-Time}
& \textbf{Temporal Repository Understanding}
& \cmark
& \cmark
& \cmark
& \cmark
& \cmark \\

\bottomrule
\end{tabular}
}

\caption{
Comparison with representative code reasoning, repository understanding, and temporal software-engineering work. \cmark / \xmark \ indicate whether a work satisfies each property; \pmark \ indicates that task queries include LLM synthesis or rewriting of real developer artifacts. The Code Understanding column identifies dedicated comprehension tasks. Sources are discussed in Section~\ref{\PaperLanguageLabelPrefix sec:related_work}.
}
\label{\PaperLanguageLabelPrefix tab:benchmark_comparison}
\end{table*}

%% file: sections/english/method.tex
\section{SWE-Time Benchmark}
\label{en:sec:method}

\afterpage{\begingroup\input{figures/figure_3}\endgroup}

\subsection{Benchmark Overview}
\label{en:sec:benchmark_overview}

SWE-Time evaluates temporal repository understanding. We model repository evolution as a commit-indexed history:
\begin{equation}
\begin{aligned}
\mathcal{H}&=(\mathcal{C},\mathcal{P},\{S_c\}_{c\in\mathcal{C}}),\\
S_c&=\operatorname{Apply}(S_{\operatorname{par}(c)},\delta_c),
\end{aligned}
\label{en:eq:repository_trajectory}
\end{equation}
where $\mathcal{H}$ denotes the repository history, $\mathcal{C}$ the commit set, $\mathcal{P}$ the Git parent relations, $S_c$ the repository state at commit $c$, and $\delta_c$ the changes introduced by $c$. The function $\operatorname{par}(c)$ identifies the parent commit, and $\operatorname{Apply}$ applies $\delta_c$ to that parent's state. The state-transition expression applies to single-parent commits, while $\mathcal{P}$ also records merge-parent relations in the Git DAG. SWE-Time assesses this ability through retrospective and prospective reasoning, combining authentic developer queries with deterministically verifiable, repository-grounded Gold. Figure~\ref{en:fig:task_formulation} summarizes task construction and evaluation.

\subsection{Design Philosophy: Repository Evolution as the Evaluation Target}
\label{en:sec:task_definition}

\paragraph{Task Backward.}
\label{en:sec:task_backward}
Task Backward is motivated by developers' need to identify the change that first introduced a reported regression. It evaluates retrospective repository understanding by asking agents to locate the first-bad commit in repository history.
The agent receives a sanitized regression report and the repository history available up to cutoff commit $\tau_i$, and must return the full 40-character SHA of the first-bad commit.
We define the Gold commit and its independently verified behavioral boundary as follows:
\begin{equation}
\begin{aligned}
c_i^\star&=\operatorname{Resolve}(e_i^{\mathrm{fb}},\mathcal{H}_{\tau_i}),\\
B_i(S_{\operatorname{par}(c_i^\star)})&=0,\qquad B_i(S_{c_i^\star})=1,
\end{aligned}
\label{en:eq:backward_boundary}
\end{equation}
where $i$ indexes cases, $\mathcal{H}_{\tau_i}$ denotes the repository history visible at cutoff commit $\tau_i$, and $e_i^{\mathrm{fb}}$ denotes developer-provided first-bad or completed bisect evidence. The function $\operatorname{Resolve}$ maps this evidence to the Gold commit $c_i^\star$ within that history. The states $S_{\operatorname{par}(c_i^\star)}$ and $S_{c_i^\star}$ are immediately before and after that commit. The independently constructed behavioral check $B_i$ returns 0 when the regression symptom is absent and 1 when it is present.

Agents may inspect and execute visible historical states and construct their own symptom checks from the report. Developer-provided first-bad evidence and independent Gold-verification checks are used only during construction and remain hidden from the agent. Each case receives a score of 1 if the predicted commit SHA exactly matches the Gold SHA, and 0 otherwise.

\paragraph{Task Forward.}
\label{en:sec:task_forward}
Before implementing a requested change, developers need to anticipate its consequences for the repository. Task Forward evaluates this prospective understanding through GitHub Actions workflow prediction. Given a developer issue, the pre-change repository state, and a fixed pull-request event and branch context, the agent predicts which GitHub Actions workflows an intended but unimplemented change would activate. The historical fixing patch and changed paths remain hidden, requiring agents to infer plausible modifications from developer intent and reason about their workflow consequences. Because reasonable implementations of the same request may yield different workflow sets, we screen candidates for workflow-level ambiguity during construction, as described in Section~\ref{en:sec:construction}. The agent prediction and Gold construction are defined as follows:
\begin{equation}
\begin{aligned}
\widehat{R}_i&=h_\theta(I_i,S_{t_i},E_i),\\
R_i^\star&=f(S_{t_i},\Delta_i^\star,E_i),
\end{aligned}
\label{en:eq:forward_gold}
\end{equation}
where $\widehat{R}_i$ denotes the predicted workflow set, $h_\theta$ the agent's prediction process parameterized by $\theta$, $I_i$ the developer issue, and $S_{t_i}$ the pre-change repository state at commit $t_i$. The context $E_i$ fixes the event type, activity type, and target branch. The Gold set $R_i^\star$ is computed by the deterministic workflow-routing function $f$ using the historical fixing PR's changed-path set $\Delta_i^\star$, which is hidden from the agent.

The routing function evaluates workflow activation from the configurations in the pre-change state:
\begin{equation}
\begin{gathered}
R_i^\star=\{w\in\mathcal{W}_i\mid\Phi(w,\Delta_i^\star,E_i)=1\},\\
\Phi(w,\Delta,E)=\phi_{\mathrm{event}}\land\phi_{\mathrm{branch}}\land\phi_{\mathrm{path}},
\end{gathered}
\label{en:eq:workflow_activation}
\end{equation}
where $\mathcal{W}_i=\mathcal{W}(S_{t_i})$ denotes the workflow configurations in the pre-change state, $w$ an individual workflow, and $\Phi$ its activation predicate for a changed-path set $\Delta$ and routing context $E$. In the second line, the right-hand terms abbreviate $\phi_{\mathrm{event}}(w,E)$, $\phi_{\mathrm{branch}}(w,E)$, and $\phi_{\mathrm{path}}(w,\Delta)$, which evaluate event type and action, target-branch filters, and changed-path include/exclude patterns with their matching precedence, respectively; $\land$ denotes logical conjunction. The conjunction summarizes the conditions for an eligible event entry; a workflow is selected if at least one eligible entry satisfies all three predicates. Substituting the historical scope $\Delta_i^\star$ and fixed context $E_i$ yields the Gold set $R_i^\star$.
Each case receives a score of 1 if the complete predicted workflow set exactly matches the Gold set, and 0 otherwise.

\begingroup\input{figures/figure_4}\endgroup

\subsection{Construction Framework: RealWorld-to-Verifiable}
\label{en:sec:construction}
\label{en:sec:evaluation_protocol}

Real software development activities do not inherently constitute verifiable evaluation tasks~\citep{kalliamvakavou2014promises}. Developer requirements are often open-ended, and their historical implementations may not be the only reasonable solutions. We introduce a \emph{RealWorld-to-Verifiable Construction Framework} that uses independently verifiable historical evidence to identify authentic development events suitable for reliable evaluation. The framework integrates target construction, Gold validation with ambiguity screening, and temporal isolation.

\subsubsection{Constructing Verifiable Targets}
We select events with traceable historical evidence and targets verifiable through repository behavior or native rules. For Task Backward, we use regression reports with explicit first-bad localization evidence, taking the reported symptoms as queries and the attributed commits as candidate Gold.

For Task Forward, we pair issues with reliably linked, merged fixing PRs, recover $S_t$ from each PR's base commit, and compute candidate workflow Gold from the historical changed paths $\Delta^*$ under fixed context $E$ using the routing function defined in \S~\ref{en:sec:task_definition}.

\subsubsection{Validating the Gold}
Historical evidence supplies candidate targets; admission additionally requires independent validation and screening for plausible alternative answers.

For Task Backward, developer localization establishes the first-bad target, while repeated symptom-derived tests or probes on the Gold commit and its parent independently verify the reported good-to-bad boundary. All admitted cases pass historical-consistency checks; conflicting, retracted, or unresolved attributions are excluded. These checks support historical localization and behavioral validity without assuming global monotonicity, as detailed in Appendix~\ref{sec:appendix_forensics_gold}.

For Task Forward, routing tests with independently specified expected outputs validate event triggers, branch and path filters, and pattern precedence. Correct routing alone does not resolve implementation ambiguity. Experienced developers review scope sufficiency, consistency between the issue and historical PR, and whether reasonable alternative implementations change workflow Gold. Candidates may be admitted with multiple reasonable implementations when the reviewed alternatives yield the same workflow Gold. Candidates with insufficient scope constraints or divergent Gold are rejected; disputed decisions undergo human adjudication, and unresolved cases are excluded. Of 1,899 candidates, 1,030 are accepted, 740 rejected, and 129 excluded as uncertain, as reported in Appendix~\ref{sec:appendix_foresight_gold}.

\begingroup\input{tables/table_3}\endgroup

\subsubsection{Temporal Isolation}
Both tasks use the complete history to construct Gold, but restrict agent access to information available at the target historical time. Backward exposes repository history up to the report-derived cutoff and a sanitized regression report, with Gold-verification probes hidden. Forward exposes only the pre-change working tree and sanitized issue, without Git history or fixing-PR evidence.

These boundaries are enforced through input sanitization, automated packaging checks of historical-view integrity and future-evidence exclusion, and runtime network restrictions. Every case also undergoes human review for indirect answer leakage. Detailed procedures are provided in Appendix~\ref{sec:appendix_historical_isolation}.

\subsubsection{Benchmark Characteristics}
SWE-Time contains 1,661 cases from 65 GitHub repositories: 631 cases from 49 repositories in Backward and 1,030 from 16 in Forward. We characterize the benchmark along two dimensions. Spatial Scope (Local/Cross) distinguishes reasoning within a local component from reasoning across repository components. Temporal Complexity (Direct/Staged) distinguishes direct target recovery from intermediate historical reasoning in Backward or modification-scope and consequence inference in Forward, as defined in Appendix~\ref{sec:appendix_historical_isolation}. Labels are assigned after admission and frozen before model evaluation. Figure~\ref{en:fig:benchmark_characteristics} summarizes repository languages, developer-event periods, and task scales.

%% file: figures/figure_3.tex
\begin{figure*}[t]
\centering
\includegraphics[width=\textwidth]{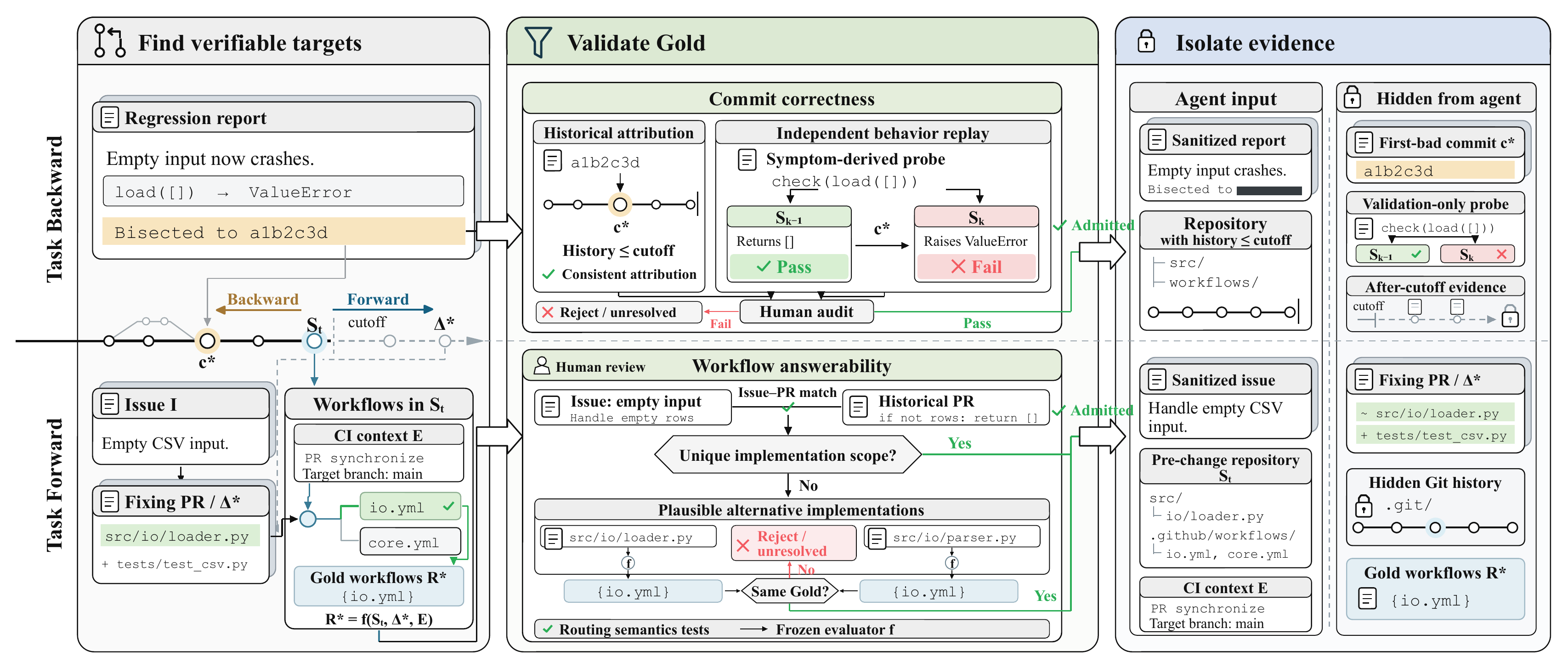}
\captionsetup{labelfont=bf}
\caption{\textbf{RealWorld-to-Verifiable Construction Framework.} Task Backward identifies the first-bad commit from a regression report and repository history up to the cutoff. Task Forward predicts activated GitHub Actions workflows from an issue and pre-change repository under a fixed CI context, with the fixing patch hidden. To instantiate these tasks, our framework establishes targets from real developer reports and issue--PR pairs. Backward validates first-bad commits through historical checks and independent behavioral replay; Forward validates workflow targets through an independently tested routing evaluator and human review of plausible implementations. Candidates with divergent Gold or unresolved judgments are excluded. Historical information boundaries separate agent-visible inputs from hidden construction evidence, with packaging checks, network isolation, and human leakage review.}
\label{\PaperLanguageLabelPrefix fig:task_formulation}
\end{figure*}

%% file: figures/figure_4.tex
\begin{figure*}[t]
\centering
\includegraphics[width=\textwidth]{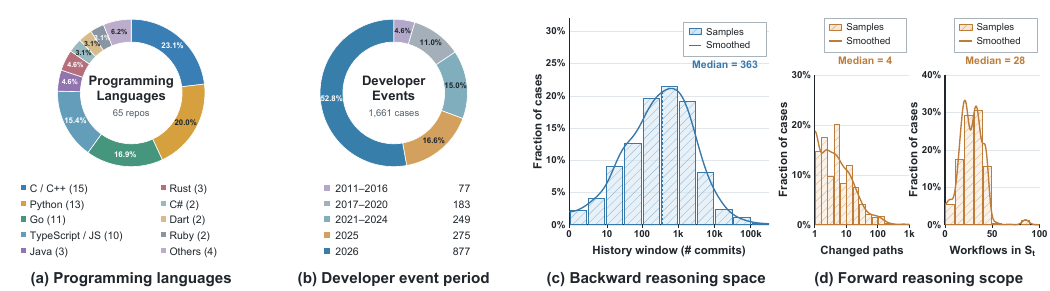}
\captionsetup{labelfont=bf}
\caption{\textbf{Characteristics of SWE-Time.} We show (a) the repository-level programming-language distribution, (b) the temporal distribution of real developer events, (c) the historical search space of Task Backward, and (d) the change and repository-context scope of Task Forward. These statistics summarize the diversity and reasoning scale of SWE-Time without relying on model-dependent measurements.}
\label{\PaperLanguageLabelPrefix fig:benchmark_characteristics}
\end{figure*}

%% file: tables/table_3.tex
\begin{table*}[!t]
\centering
\definecolor{MainOverallGray}{gray}{0.96}
\definecolor{MainAverageGray}{gray}{0.91}
\newcommand{\MainAverageScore}[1]{{%
  \setbox0=\hbox{\textbf{Overall}}%
  \makebox[0pt][c]{\color{MainAverageGray}%
    \rule[-\dp\strutbox]{\dimexpr\wd0+2\tabcolsep\relax}{\dimexpr\ht\strutbox+\dp\strutbox\relax}}%
  \makebox[0pt][c]{#1}%
}}
\resizebox{\textwidth}{!}{
\begin{tabular}{@{}l*{18}{C}@{}}
\toprule

\multirow{3}{*}{\textbf{Model}}
& \multicolumn{6}{c}{\textbf{Task \textit{Backward}}}
& \multicolumn{6}{c}{\textbf{Task \textit{Forward}}}
& \multicolumn{6}{c}{\multirow{3}{*}{\textbf{\textit{SWE-Time} Avg.}}} \\

\cmidrule(lr){2-7}
\cmidrule(lr){8-13}

&
\multicolumn{2}{c}{\textbf{Direct}}
& \multicolumn{2}{c}{\textbf{Staged}}
& \multicolumn{2}{c}{\multirow{2}{*}{\textbf{Overall}}}

& \multicolumn{2}{c}{\textbf{Direct}}
& \multicolumn{2}{c}{\textbf{Staged}}
& \multicolumn{2}{c}{\multirow{2}{*}{\textbf{Overall}}}

& \multicolumn{6}{c}{} \\

\cmidrule(lr){2-3}
\cmidrule(lr){4-5}
\cmidrule(lr){8-9}
\cmidrule(lr){10-11}

& \textbf{Local}
& \textbf{Cross}
& \textbf{Local}
& \textbf{Cross}
& \multicolumn{2}{c}{}

& \textbf{Local}
& \textbf{Cross}
& \textbf{Local}
& \textbf{Cross}
& \multicolumn{2}{c}{}

& \multicolumn{6}{c}{} \\

\midrule

\multicolumn{19}{c}{\textbf{Claude Code}} \\
\midrule

Claude Opus 4.8
& \textbf{77.7} & \textbf{63.0} & \textbf{61.0} & 45.5
& \multicolumn{2}{c}{\cellcolor{MainOverallGray}\textbf{61.2}}
& \textbf{81.1} & \textbf{78.5} & 56.3 & \underline{58.0}
& \multicolumn{2}{c}{\cellcolor{MainOverallGray}\underline{62.7}}
& \multicolumn{6}{c}{\MainAverageScore{\textbf{61.9}}} \\

Claude Sonnet 4.6
& 63.1 & 47.4 & 47.7 & 36.5
& \multicolumn{2}{c}{\cellcolor{MainOverallGray}48.0}
& 68.9 & 61.0 & 46.5 & 43.8
& \multicolumn{2}{c}{\cellcolor{MainOverallGray}49.0}
& \multicolumn{6}{c}{\MainAverageScore{48.5}} \\

Gemini 3.1 Pro
& 50.8 & 37.0 & 39.0 & 20.5
& \multicolumn{2}{c}{\cellcolor{MainOverallGray}36.3}
& 37.8 & 23.8 & 20.0 & 15.5
& \multicolumn{2}{c}{\cellcolor{MainOverallGray}19.4}
& \multicolumn{6}{c}{\MainAverageScore{27.9}} \\

Gemini 3.8 Flash
& 64.6 & 58.4 & 56.4 & 41.7
& \multicolumn{2}{c}{\cellcolor{MainOverallGray}55.0}
& 74.3 & 69.2 & 52.6 & 47.3
& \multicolumn{2}{c}{\cellcolor{MainOverallGray}54.0}
& \multicolumn{6}{c}{\MainAverageScore{54.5}} \\

Gemini 3.6 Flash
& 59.2 & 48.0 & 43.6 & 31.4
& \multicolumn{2}{c}{\cellcolor{MainOverallGray}45.0}
& 58.1 & 54.7 & 40.5 & 38.5
& \multicolumn{2}{c}{\cellcolor{MainOverallGray}43.0}
& \multicolumn{6}{c}{\MainAverageScore{44.0}} \\

DeepSeek V4 Flash
& 55.4 & 48.0 & 48.8 & 31.4
& \multicolumn{2}{c}{\cellcolor{MainOverallGray}45.6}
& 58.1 & 57.6 & 39.1 & 43.9
& \multicolumn{2}{c}{\cellcolor{MainOverallGray}46.2}
& \multicolumn{6}{c}{\MainAverageScore{45.9}} \\

GLM 5.3
& 69.2 & 57.8 & \underline{59.3} & \textbf{47.4}
& \multicolumn{2}{c}{\cellcolor{MainOverallGray}58.0}
& \underline{78.4} & \underline{77.3} & \textbf{61.4} & 54.3
& \multicolumn{2}{c}{\cellcolor{MainOverallGray}61.4}
& \multicolumn{6}{c}{\MainAverageScore{59.7}} \\

GLM 5.2
& 63.8 & 53.2 & 57.0 & 42.3
& \multicolumn{2}{c}{\cellcolor{MainOverallGray}53.7}
& 74.3 & 73.3 & 54.4 & 53.3
& \multicolumn{2}{c}{\cellcolor{MainOverallGray}58.3}
& \multicolumn{6}{c}{\MainAverageScore{56.0}} \\

Qwen 3.8 Max
& 69.2 & 59.0 & 58.1 & \underline{46.8}
& \multicolumn{2}{c}{\cellcolor{MainOverallGray}57.8}
& 77.0 & 76.2 & \underline{57.7} & \textbf{60.8}
& \multicolumn{2}{c}{\cellcolor{MainOverallGray}\textbf{63.9}}
& \multicolumn{6}{c}{\MainAverageScore{\underline{60.9}}} \\

Qwen 3.6 Plus
& 56.9 & 45.1 & 44.8 & 28.8
& \multicolumn{2}{c}{\cellcolor{MainOverallGray}43.4}
& 37.8 & 17.4 & 19.1 & 16.0
& \multicolumn{2}{c}{\cellcolor{MainOverallGray}18.4}
& \multicolumn{6}{c}{\MainAverageScore{30.9}} \\

Kimi K3
& \underline{73.1} & \underline{62.4} & 56.4 & 44.2
& \multicolumn{2}{c}{\cellcolor{MainOverallGray}\underline{58.5}}
& 75.7 & 75.6 & 54.9 & 57.8
& \multicolumn{2}{c}{\cellcolor{MainOverallGray}61.5}
& \multicolumn{6}{c}{\MainAverageScore{60.0}} \\

Kimi K2.6
& 66.9 & 54.3 & 55.8 & 41.0
& \multicolumn{2}{c}{\cellcolor{MainOverallGray}54.0}
& 74.3 & 73.3 & 54.0 & 51.0
& \multicolumn{2}{c}{\cellcolor{MainOverallGray}57.0}
& \multicolumn{6}{c}{\MainAverageScore{55.5}} \\

Hunyuan 3
& 31.5 & 25.4 & 23.8 & 14.1
& \multicolumn{2}{c}{\cellcolor{MainOverallGray}23.5}
& 36.5 & 27.3 & 25.6 & 22.7
& \multicolumn{2}{c}{\cellcolor{MainOverallGray}25.0}
& \multicolumn{6}{c}{\MainAverageScore{24.3}} \\
\midrule
\multicolumn{19}{c}{\textbf{Codex}} \\
\midrule

GPT 5.5
& 73.8 & 65.3 & 61.0 & 43.6
& \multicolumn{2}{c}{\cellcolor{MainOverallGray}60.5}
& 77.0 & 79.7 & 59.1 & 58.3
& \multicolumn{2}{c}{\cellcolor{MainOverallGray}63.4}
& \multicolumn{6}{c}{\MainAverageScore{62.0}} \\

GPT 5.4
& 69.2 & 59.5 & 58.7 & 36.5
& \multicolumn{2}{c}{\cellcolor{MainOverallGray}55.6}
& 74.3 & 77.3 & 54.4 & 58.0
& \multicolumn{2}{c}{\cellcolor{MainOverallGray}61.7}
& \multicolumn{6}{c}{\MainAverageScore{58.7}} \\

\bottomrule
\end{tabular}
}

\caption{
Performance (\%) on Tasks Backward and Forward.
The former traces an observed behavior back to its historical cause, while the latter reasons from a change intent to its repository-level consequences.
Claude Code is the main evaluation harness; Codex provides supplementary references for harness diversity.
We report \textbf{temporal complexity} (\textbf{\textit{Direct}} / \textbf{\textit{Staged}}) using task-specific criteria: historical-localization difficulty in Backward and direct routing versus additional change-scope or impact-propagation inference in Forward.
We also report \textbf{spatial scope} (\textbf{\textit{Local}}: local scope; \textbf{\textit{Cross}}: across files, components, or workflows).
\textbf{Overall} aggregates each task, and \textbf{\textit{SWE-Time} Avg.} averages the two task-level scores.
}
\label{\PaperLanguageLabelPrefix tab:combined_results}
\end{table*}

%% file: sections/english/experiments.tex
\section{Experiments}
\label{en:sec:experiments}

We organize our experiments around four research questions. \textbf{\S\ref{en:sec:main_results}} establishes the capabilities and limitations of current coding agents on SWE-Time. \textbf{\S\ref{en:sec:temporal_evidence_analysis}} examines dependence on historical code evidence and sensitivity to repository-state changes. \textbf{\S\ref{en:sec:developer_intent_analysis}} examines how developer intent and gold change-scope information contribute to performance. \textbf{\S\ref{en:sec:complexity_analysis}} relates performance to history-window, changed-path, and workflow-context sizes.

\subsection{Experimental Setup}
\label{en:sec:experimental_setup}

We evaluate 15 coding agents: 13 under Claude Code~\citep{anthropic2026claudecode} for the main model comparison, and GPT 5.5 and GPT 5.4 under Codex~\citep{openai2026codex} as supplementary references to broaden harness coverage. Within each harness, models share the tool interface, task instructions, and inference-budget policy. Model rankings are reported within harness groups. All evaluations enforce the historical information boundaries in Section~\ref{en:sec:evaluation_protocol}; execution details appear in Appendix~\ref{sec:appendix_historical_isolation}.

RQ1 baselines and RQ2--RQ3 use fixed 300-case diagnostic subsets per task. Claude Opus 4.8, the highest-scoring Claude Code agent by SWE-Time Avg., serves as the reference. Controlled conditions reuse the same cases; Overall is case-weighted on each subset. RQ4 analyzes this reference agent's full-task predictions.

\subsection{RQ1: Overall Capability}
\label{en:sec:main_results}

Table~\ref{en:tab:combined_results} reports coding-agent performance on the full SWE-Time benchmark. Claude Opus 4.8 leads the Claude Code group with a SWE-Time Avg. of 61.9, scoring 61.2 on Backward and 62.7 on Forward. Qwen 3.8 Max leads Forward within this harness at 63.9. These results show substantial headroom on SWE-Time.

For every agent, Direct cases outperform Staged cases in both tasks, indicating greater difficulty with intermediate historical localization or prospective scope and consequence inference. Local--Cross differences are less consistent.

We compare recency, retrieval, and repository-prior baselines with Claude Opus 4.8 on the same diagnostic cases, as reported in Table~\ref{en:tab:shortcut_baselines}. For Backward, Recent-1 returns the latest commit in the visible candidate history. The two BM25 baselines~\citep{robertson2009bm25} rank visible commits using either their commit messages or changed paths and unified diffs, returning the top-ranked commit for the sanitized regression report. Repository Prior evaluates $f(S_t,\{p\},E)$ for every Git-tracked path $p$ and predicts the most frequent complete workflow set, without the issue or $\Delta^*$.

Under the same historical information boundary, our task-specific Backward baseline, History Retrieval + LLM Reranking, retrieves visible candidate commits and uses Claude Opus 4.8 to rerank them against the regression report. Fix-conditioned localization methods such as AgentSZZ and LLMBisect~\citep{lyu2026agentszz,zhang2026llmbisect} require fixing evidence excluded by our historical cutoff, preventing direct evaluation under their original input assumptions.

\begingroup\input{tables/table_4}\endgroup

As shown in Table~\ref{en:tab:shortcut_baselines}, lightweight baselines remain substantially below the general agent. On Backward, LLM reranking reaches 14.0\%, improving on BM25's 9.0\% but remaining below the agent's 62.0\%. On Forward, Repository Prior reaches 31.3\%, compared with 63.3\%. These results indicate that retrieval and repository-level priors alone are insufficient for the two tasks.

The repository-level robustness analysis reported in Appendix~\ref{sec:appendix_repository_robustness} finds case-weighted and repository-macro scores within 2.4 percentage points across five Forward configurations, with the ordering of the four Claude Code models unchanged after removing any single repository.

\subsection{RQ2: Temporal Repository Evolution}
\label{en:sec:temporal_evidence_analysis}

We examine whether performance depends on repository evolution through two controlled interventions summarized in Table~\ref{en:tab:temporal_sensitivity}. In Backward, we remove historical code states and diffs while preserving the regression report, commit sequence, and metadata. In Forward, we hold developer intent and change scope fixed while placing the same change in a different historical repository state.

\begingroup\input{tables/table_5}\endgroup

Overall falls from 62.0 to 28.0, a 34.0-point drop, with declines in all four subsets. Historical code evidence thus contributes substantially to first-bad localization beyond commit metadata.

For Forward, we pair each original state with the nearest valid earlier state in the same first-parent history that changes workflow Gold and reverses at least one shared workflow's activation. We hold the developer issue, change scope ($\Delta^*$), and evaluation context ($E$) fixed, excluding semantically invalid pairs before evaluation. Pair Exact requires correctness in both states; Invariance Error measures unchanged predictions despite different Gold targets among scorable pairs; Invalid denotes pairs containing unscorable outputs.

Table~\ref{en:tab:temporal_sensitivity}(b) shows that Exact drops from 63.3\% to 54.3\%, with 48.7\% Pair Exact, 12.8\% Invariance Error, and 6.0\% Invalid. Fewer than half of pairs are solved in both states, showing that prediction changes do not ensure correct adaptation to historical routing transitions.

\subsection{RQ3: Developer-Intent Contribution}
\label{en:sec:developer_intent_analysis}

Using the interventions shown in Figure~\ref{en:fig:developer_intent_grounding}, we examine the contributions of developer-query information and gold change scope while holding repository context fixed. In Backward, symptom ablation reduces Overall from 62.0 to 41.7, a 20.3-point drop, with declines in all four subsets. Regression symptoms thus provide substantial information for first-bad localization.

\begingroup\input{figures/figure_5}\endgroup

In Forward, removing the issue lowers Overall from 63.3 to 37.3; adding gold change scope to Full Issue raises it to 85.7. The 26.0-point Full-vs.-No-Issue gap measures the contribution of developer intent, whereas the additional 22.3-point Oracle gain suggests that change-scope inference is a substantial source of difficulty. This gain is especially large on Staged cases; even with the gold scope, residual errors remain in consequence reasoning.

\subsection{RQ4: Difficulty Analysis}
\label{en:sec:complexity_analysis}

As shown in Figure~\ref{en:fig:performance_complexity}, performance decreases as the reasoning space grows. In Backward, performance drops from 65.2 to 57.0 across history-window bins. Forward shows larger declines: 21.0 percentage points across changed-path bins and 31.4 across workflow-context bins. The largest gap occurs with workflow context, suggesting that prospective consequence reasoning becomes particularly difficult as the relevant repository context expands. Overall, larger historical search spaces in Backward and more complex pre-change repository contexts in Forward are associated with lower performance.

\begingroup\input{figures/figure_6}\endgroup

%% file: tables/table_4.tex
\begin{table}[!ht]
\centering
\captionsetup{labelfont=bf}
\begingroup
\small
\definecolor{ExperimentReferenceGray}{gray}{0.95}
\setlength{\heavyrulewidth}{0.6pt}
\setlength{\lightrulewidth}{0.35pt}
\setlength{\tabcolsep}{0pt}
\renewcommand{\arraystretch}{1.08}
\begin{tabular}{@{}>{\raggedright\arraybackslash}p{0.23\columnwidth}>{\raggedright\arraybackslash}p{0.58\columnwidth}>{\raggedleft\arraybackslash}p{\dimexpr\columnwidth-0.23\columnwidth-0.58\columnwidth\relax}@{}}
\toprule
\textbf{Task} & \textbf{Baseline / Agent} & \textbf{Overall (\%)} \\
\midrule
\textbf{Backward} & Recent-1 & 0.7 \\
& BM25 (commit message) & 2.3 \\
& BM25 (diff / paths) & 9.0 \\
& History Retrieval + LLM Reranking & 14.0 \\
\rowcolor{ExperimentReferenceGray}[0pt][0pt]
& Claude Opus 4.8 (General Agent) & 62.0 \\
\midrule
\textbf{Forward} & Repository Prior & 31.3 \\
\rowcolor{ExperimentReferenceGray}[0pt][0pt]
& Claude Opus 4.8 (General Agent) & 63.3 \\
\bottomrule
\end{tabular}
\endgroup
\caption{\textbf{Baseline and shortcut analysis.}}
\label{\PaperLanguageLabelPrefix tab:shortcut_baselines}
\end{table}

%% file: tables/table_5.tex
\begin{table}[!ht]
\centering
\captionsetup{labelfont=bf}
\begingroup
\small
\definecolor{SWEHistoryGain}{HTML}{4C78A8}
\definecolor{ExperimentScoreGray}{gray}{0.95}
\newcommand{\HistoryGain}[1]{\raisebox{-0.5ex}{\fontsize{7}{8}\selectfont\textcolor{SWEHistoryGain}{+#1}}}
\newcommand{\ScoreHeading}[1]{\makebox[1.75em][c]{\textbf{#1}}\hphantom{\HistoryGain{00.0}}}
\newcommand{\ScoreWithoutGain}[1]{#1\hphantom{\HistoryGain{00.0}}}
\setlength{\tabcolsep}{1.6pt}
\setlength{\heavyrulewidth}{0.6pt}
\setlength{\lightrulewidth}{0.35pt}
\renewcommand{\arraystretch}{1.08}
\begin{tabular}{@{}p{\columnwidth}@{}}
\toprule
\textbf{Task Backward}\par\vspace{2pt}
\noindent\begin{tabular*}{\linewidth}[t]{@{\extracolsep{\fill}}lrrrr>{\columncolor{ExperimentScoreGray}[\tabcolsep][0pt]}r@{}}
\textbf{Condition} & \ScoreHeading{D-L} & \ScoreHeading{D-C}
& \ScoreHeading{S-L} & \ScoreHeading{S-C} & \ScoreHeading{O} \\
\midrule
Meta. only & \ScoreWithoutGain{47.8} & \ScoreWithoutGain{24.7}
& \ScoreWithoutGain{23.1} & \ScoreWithoutGain{18.1} & \ScoreWithoutGain{28.0} \\
Full hist. & 82.6\HistoryGain{34.8} & 66.7\HistoryGain{42.0}
& 57.7\HistoryGain{34.6} & 41.7\HistoryGain{23.6}
& 62.0\HistoryGain{34.0} \\
\end{tabular*}\\
\midrule
\textbf{Task Forward}\par\vspace{2pt}
\noindent\begin{tabular*}{\linewidth}[t]{@{\extracolsep{\fill}}cccc>{\columncolor{ExperimentScoreGray}[\tabcolsep][0pt]}c@{}}
\textbf{Original} & \textbf{Paired} & \textbf{Inv. Err.} $\downarrow$ & \textbf{Invalid} $\downarrow$ & \textbf{Pair Exact} \\
\midrule
63.3 & 54.3 & 12.8 & 6.0 & 48.7 \\
\end{tabular*}\\
\bottomrule
\end{tabular}
\endgroup
\caption{\textbf{Temporal repository evolution.} Backward removes historical code while retaining metadata; Forward contrasts historical states with issue, change scope, and routing context fixed. Task scores and paired-state metrics are percentages. Blue subscripts report gains over Metadata-Only.}
\label{\PaperLanguageLabelPrefix tab:temporal_sensitivity}
\end{table}

%% file: figures/figure_5.tex
\begin{figure}[!ht]
\centering
\includegraphics[width=\columnwidth]{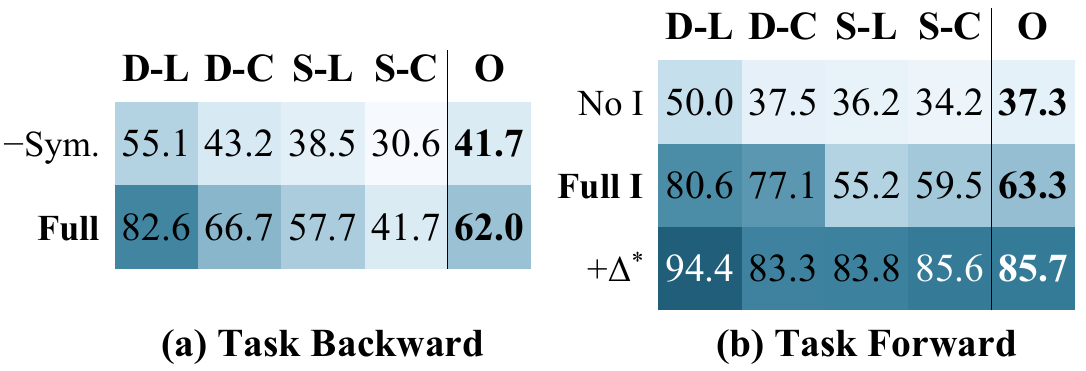}
\captionsetup{labelfont=bf}
\caption{\textbf{Developer-intent contribution.} We intervene on the developer query while holding repository context fixed. $-$Sym. removes the regression symptom, and $+\Delta^*$ provides the gold change scope.}
\label{\PaperLanguageLabelPrefix fig:developer_intent_grounding}
\end{figure}

%% file: figures/figure_6.tex
\begin{figure}[!ht]
\centering
\includegraphics[width=\columnwidth]{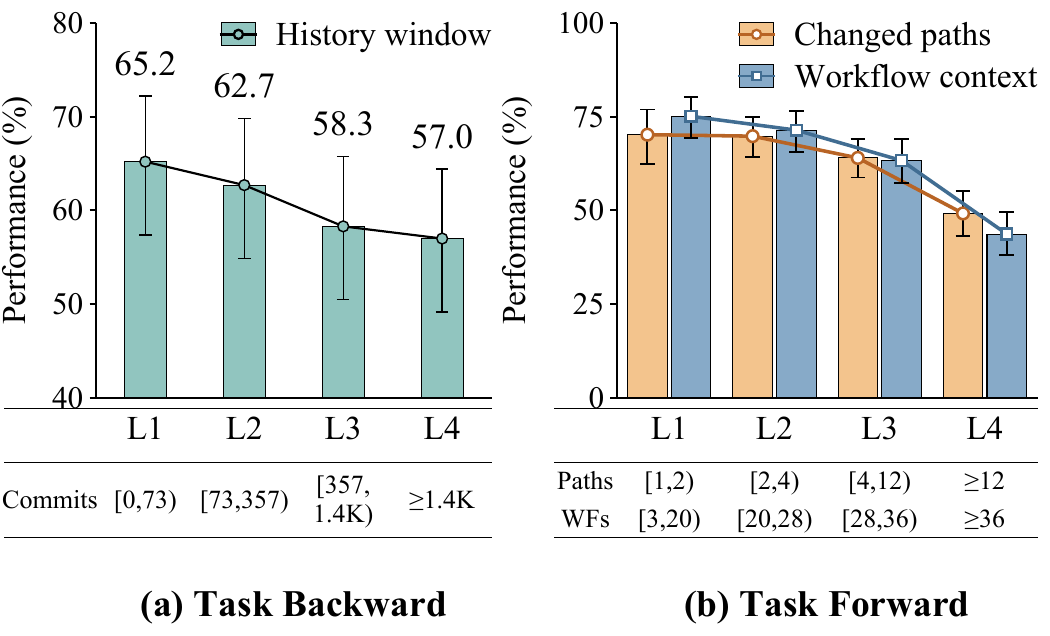}
\captionsetup{labelfont=bf}
\caption{\textbf{Performance across task-complexity factors.} We group cases by (a) searchable history window in Task Backward and (b) changed-path scope and workflow-context size in Task Forward. Bars show performance within each bin, and error bars denote 95\% Wilson confidence intervals~\citep{wilson1927probable}. Tables below the panels report the corresponding bin ranges.}
\label{\PaperLanguageLabelPrefix fig:performance_complexity}
\end{figure}

%% file: sections/english/conclusion.tex
\section{Conclusion}
\label{en:sec:conclusion}

We introduced SWE-Time, extending repository understanding from static snapshots to evolving trajectories. SWE-Time contains 1,661 cases from 65 GitHub repositories, evaluating first-bad localization and prospective prediction of GitHub Actions workflow activation. Our RealWorld-to-Verifiable Construction Framework pairs authentic developer queries with independently validated Gold for deterministic evaluation. Evaluation of 15 coding agents under two harnesses reveals substantial headroom: Claude Opus 4.8 leads the Claude Code group with a SWE-Time Avg. of 61.9. Controlled analyses further show that historical code evidence and developer intent both contribute materially to performance, while maintaining correct workflow predictions across paired repository states remains difficult. We hope SWE-Time provides a foundation for developing coding agents that understand not only repository state, but also repository evolution.

%% file: sections/english/limitations.tex
\section*{Limitations}

SWE-Time focuses on two concrete forms of temporal repository understanding: retrospective localization of first-bad commits and prospective prediction of GitHub Actions workflow activation. These tasks capture complementary directions along repository history, but they do not cover the full range of temporal reasoning required in software engineering. In particular, Task Forward evaluates CI routing under a fixed pull-request event context, so generalization to other CI systems, build pipelines, or non-GitHub development environments remains to be studied. The benchmark also relies on repositories with sufficiently recoverable public issue, pull-request, commit, and workflow histories, which may bias the dataset toward projects with richer development records. Because the repositories are public, models may have encountered their code or development records during pretraining.

Our experiments cover 15 coding agents under two agent harnesses. Although they span several recent model families, the reported results should not be assumed to characterize future models or alternative agent frameworks.

%% file: sections/appendix_case_cards.tex
\newenvironment{SWETimeCaseCard}[3]{%
  \begin{tcolorbox}[%
    breakable,
    colback=black!5,
    colframe=black!5,
    boxrule=0pt,
    arc=0pt,
    left=6pt,
    right=6pt,
    top=6pt,
    bottom=6pt,
    before skip=6pt,
    after skip=5pt,
    before upper={%
      \normalsize\raggedright
      \setlength{\parindent}{0pt}%
      \setlength{\parskip}{3pt}%
      \textbf{#1 \textbullet{} #2}\hfill\textbf{#3}\par
      \nobreak\smallskip\nobreak
      \let\SWETimeCaseSavedPar\par
      \def\par{\let\par\SWETimeCaseSavedPar\ifhmode\SWETimeCaseSavedPar\fi}%
    }
  ]%
}{%
  \end{tcolorbox}%
}
\newenvironment{SWETimeCaseOutputs}{%
  \par\setlength{\tabcolsep}{3pt}%
  \renewcommand{\arraystretch}{1.07}%
  \begin{tabular}{@{}>{\raggedright\arraybackslash}p{\dimexpr\linewidth-58pt\relax}>{\centering\arraybackslash}p{52pt}@{}}%
  \toprule\textbf{Reviewed scope} & \textbf{Output}\\\midrule
}{\bottomrule\end{tabular}\par}
\newcommand{\SWETimeCaseSource}[2]{%
  \par\smallskip{\footnotesize Historical PR \##1. Case ID: \path{#2}.\par}%
}

%% file: sections/english/appendix_a.tex
\setcounter{table}{0}
\setcounter{equation}{0}
\renewcommand{\thetable}{A.\arabic{table}}
\renewcommand{\theequation}{A.\arabic{equation}}
\renewcommand{\theHtable}{appendix.A.\arabic{table}}
\renewcommand{\theHequation}{appendix.A.\arabic{equation}}
\section{First-Bad Gold Attribution and Validation}
\label{sec:appendix_forensics_gold}

Backward evaluates whether an agent can trace a reported regression to the commit that first introduced it within the available repository history. Its Gold originates from a developer's completed historical localization of the reported regression. We subsequently resolve and validate the attributed commit, screen cases for ambiguity and information leakage, and independently check the behavioral transition at the attributed boundary. This appendix describes the construction evidence for the 631 admitted cases from 49 repositories and illustrates why some seemingly relevant reports are excluded.

\subsection{First-Bad Gold Attribution}

\paragraph{Developer-grounded origin.}
We discover real GitHub regression reports through high-precision references to historical localization, such as \emph{first bad commit}, \emph{bisected to}, \emph{regression introduced by}, and \emph{bisect found}. A candidate must contain a completed developer attribution that identifies a specific commit as the first bad revision for the reported symptom. A commit range, a tentative suspect, or an unsubstantiated mention of a SHA is insufficient. Both direct first-bad statements and completed bisect reports are eligible; their differing linguistic formulations do not change the target definition.

\paragraph{Resolving the historical target.}
We associate each attribution with its original issue or comment and resolve abbreviated hashes against the source repository to a unique, full 40-character commit SHA. We recover a historical cutoff at which the regression report is available and verify that the attributed commit belongs to the permitted history. In particular, the target must be reachable from the cutoff and satisfy the benchmark's default-branch and non-merge selection constraints.

\paragraph{Selection.}
Semantic screening retained 666 candidates, of which 657 were materialized for evaluation. Final human review retained 631 cases. Table~\ref{tab:appendix_forensics_selection} summarizes these successive selection stages.

\begin{table}[!ht]
\centering
\begingroup
\small
\setlength{\heavyrulewidth}{0.6pt}
\setlength{\lightrulewidth}{0.35pt}
\setlength{\tabcolsep}{0pt}
\renewcommand{\arraystretch}{1.08}
\begin{tabular}{@{}>{\raggedright\arraybackslash}p{0.80\columnwidth}>{\raggedleft\arraybackslash}p{\dimexpr\columnwidth-0.80\columnwidth\relax}@{}}
\toprule
\textbf{Selection stage} & \textbf{Cases} \\
\midrule
Accepted by semantic screening & 666 \\
Materialized for evaluation & 657 \\
Retained after final human review & 631 \\
\bottomrule
\end{tabular}
\endgroup
\caption{Backward candidate selection after semantic screening.}
\label{tab:appendix_forensics_selection}
\end{table}

The resulting cases retain the regression symptom as the developer-facing query but remove the attribution used to establish the Gold. Thus, the agent must recover the historical origin from the report and the repository, rather than read the bisect answer from the input.

\subsection{Historical and Behavioral Validation}

Gold validation comprises historical attribution, structural consistency, and independent behavioral confirmation.

\paragraph{Historical and structural checks.}
For each candidate, we verify the identity and location of the attributed commit in the Git history, its relationship to the cutoff, and the consistency of its parent information. Table~\ref{tab:appendix_forensics_structural} reports the results for the admitted set. Candidates with Gold commits that were merge commits, outside the permitted default-branch history, or later than the cutoff were excluded during construction.

\begin{table}[!ht]
\centering
\begingroup
\small
\setlength{\heavyrulewidth}{0.6pt}
\setlength{\lightrulewidth}{0.35pt}
\setlength{\tabcolsep}{0pt}
\renewcommand{\arraystretch}{1.08}
\begin{tabular}{@{}>{\raggedright\arraybackslash}p{0.80\columnwidth}>{\raggedleft\arraybackslash}p{\dimexpr\columnwidth-0.80\columnwidth\relax}@{}}
\toprule
\textbf{Validation check} & \textbf{Passed} \\
\midrule
Full, valid 40-character Gold SHA & 631 \\
Gold reachable from the historical cutoff & 631 \\
Gold is a non-merge commit & 631 \\
Parent information consistent across repository checks & 631 \\
Cutoff and temporal-consistency validation & 631 \\
Final construction validation gate & 631 \\
\bottomrule
\end{tabular}
\endgroup
\caption{Recorded structural validation of the 631 admitted Backward cases.}
\label{tab:appendix_forensics_structural}
\end{table}

\paragraph{Independent behavioral confirmation.}
During dataset construction, the 631 admitted cases underwent independent good-to-bad verification in a separate, build-capable execution environment. For each case, the probe was defined from two observable parts of the developer's regression report: the \emph{trigger} (reproduction steps, input, or target test) and the \emph{reported failure} (such as a specific diagnostic, crash, assertion, or failing test). Its pass/fail criterion was fixed for that case before comparing revisions and evaluated through machine-checkable signals, including exit status, test outcomes, crash signals, or diagnostic substrings. A nonzero exit code counted as evidence of the regression only when it corresponded to the reported failure, rather than an unrelated build or environment error~\citep{zhu2023reproducibility}.

For the same trigger and symptom-specific predicate $B_i$, we classified a revision as bad if the reported failure appeared, and good if the trigger completed without that failure (for example, when the relevant test passed). We applied the same criterion under comparable conditions to the Gold commit $c_i^\star$ and its immediate parent $\operatorname{par}(c_i^\star)$, verifying the local good-to-bad boundary defined in Equation~\ref{en:eq:backward_boundary}.

Each revision was probed repeatedly, and a stable classification required agreement across repeated executions~\citep{luo2014flaky,gruber2021flaky}. Inconsistent outcomes were treated as flaky; failures to exercise the trigger or reproduce the specific reported symptom were treated as non-reproducible rather than counted as evidence of a bad revision. These checks used the independent verification environment, not the repository snapshot distributed to agents, which intentionally did not require project build tools.

Developer localization establishes the historical first-bad target, while repeated adjacent-state probes independently confirm its behavioral boundary. Historical checks and ambiguity screening corroborate the attribution for the reported symptom.

\subsection{Ambiguity Screening and Human Review}

A developer's reference to a commit is not sufficient for admission unless it clearly identifies the origin of the same regression symptom presented to the agent. We therefore screen for three classes of problems.

\paragraph{Attribution ambiguity.}
Candidates are flagged when the issue contains competing culprit commits, a later correction or retraction, an explicitly uncertain assertion, or an attribution that refers to another bug. An unresolved conflict cannot be converted into a unique Gold by choosing the most recent or most frequently mentioned SHA.

\paragraph{Insufficient developer-facing evidence.}
After removing answer-bearing material, the remaining query must still describe an observable regression clearly enough to support meaningful historical investigation. Reports consisting only of version metadata, commit lists, or an isolated bisect transcript are not admitted if their actual symptoms or reproduction context are missing.

\paragraph{Historical inconsistency and leakage.}
We check that the attributed revision is compatible with the historical cutoff and selected branch, and that the sanitized report does not retain a revealing SHA, culprit link, bisect outcome, or equivalent answer-bearing clue. Sanitization removes such material rather than rewriting the developer's technical symptom. The historical information boundary and evaluation-package isolation are described separately in Appendix~\ref{sec:appendix_historical_isolation}.

Semantic screening uses a five-part rubric covering attribution credibility, authentic regression symptoms, avoidance of answer-derived evidence, absence of answer leakage, and sufficient information. An LLM-assisted pass proposes decisions supported by quotations that are checked against the original source material; Gold SHAs remain grounded in developer attributions. The review process additionally uses an 80-case human calibration sample, explicit human rulings for disputed candidates, and a human review ledger. The final review excluded 26 further cases from the 657 materialized candidates.

Admission requires evidence supporting a historical first-bad target for the same regression symptom. The examples below illustrate both accepted attributions and rejected candidates.

\subsection{Worked Regression Cases}

\label{sec:appendix_forensics_cases}

The cases below are drawn from the construction records. For accepted cases, the developer's bisect attribution is used to determine the Gold, while the agent-visible query excludes the attribution itself. Rejected cases illustrate where the same construction procedure refuses to turn a historical SHA into a benchmark answer.

\begin{SWETimeCaseCard}{A1}{curl-9931}{Accept}

\emph{An explicit bisect across a connection-layer refactor.}

\textbf{Input --- regression symptom.} An HTTP/3/QUIC request to \path{127.0.0.1:8443} fails with \texttt{curl: (35) error:0A00010B:SSL routines::wrong version number}. The sanitized report retains the connection mode, error, and diagnostic context.

\textbf{Historical evidence.} A developer explicitly identifies the Gold revision as \emph{the first bad commit, according to my bisect}. This attribution is removed from the agent-facing report. The searchable history window contains 26 commits.

\textbf{Repository transition.} The Gold, \emph{lib: connection filters (cfilter) addition}, is a 49-file connection-layer refactor spanning \path{lib/cfilters.*}, \path{connect.c}, \path{url.c}, and TLS/SSH components. Localization requires relating the connection failure to this change across components.

\textbf{Validation and decision.} The completed bisect establishes first-bad attribution; historical and structural checks pass, and independent parent/Gold probes confirm the behavioral boundary. \textbf{Accept:} a specific symptom, definite attribution, and valid historical target.

\par\smallskip{\footnotesize Gold SHA: \path{dafdb20a26d0c890e83dea61a104b75408481ebd}.\par

Case ID: \path{curl_curl__9931__85aaefd8}.\par

Source: \href{https://github.com/curl/curl/issues/9931}{curl/curl \#9931}.\par}

\end{SWETimeCaseCard}

\begin{SWETimeCaseCard}{A2}{flutter-177420}{Accept}

\emph{A regression introduced by an engine-roll commit.}

\textbf{Input --- regression symptom.} On an iOS simulator, \path{BackdropFilterLayer} allows blur behind a \path{CupertinoSheetRoute} to extend beyond its rounded boundary after the animation settles. Sanitization removes the bisect attribution and answer-bearing reference.

\textbf{Historical evidence.} Completed developer localization identifies the Gold revision within a searchable history window of 43,902 commits.

\textbf{Repository transition.} An automated Flutter Engine roll changes only \path{bin/internal/engine.version}, updating the engine consumed by Flutter. The first-bad target is this repository revision, rather than a lower-level engine commit outside its history.

\textbf{Validation and decision.} The completed bisect identifies the repository-level transition, which passes the structural and behavioral checks. \textbf{Accept:} an automated version-roll commit marking the reported regression's historical boundary.

\par\smallskip{\footnotesize Gold SHA: \path{7bb9352911d0471568a77fc57b94a646fe9571af}.\par

Case ID: \path{flutter_flutter__177420__bfb00274}.\par

Source: \href{https://github.com/flutter/flutter/issues/177420}{flutter/flutter \#177420}.\par}

\end{SWETimeCaseCard}

\begin{SWETimeCaseCard}{R1}{curl-17030}{Reject}

\emph{Symptom--Attribution Mismatch.}

\textbf{Input --- regression symptom.} Configuration lines of certain lengths are ignored.

\textbf{Historical evidence.} Extraction selected the candidate revision listed below from a bisect discussion of a distinct \path{netrc}-parsing problem. The configuration-line regression was attributed to a revision beginning \path{d8618f4d}. These attributions concern different symptoms.

\textbf{Validation and decision.} \textbf{Reject:} the extracted bisect identifies another problem and does not establish a first-bad target for the agent-visible symptom. A valid SHA alone does not resolve this mismatch.

\par\smallskip{\footnotesize Candidate SHA: \path{142ac257b3242459b284020c59f1902b9687a954}.\par

Candidate ID: \path{curl_curl__17030__6be98a4c}.\par

Source: \href{https://github.com/curl/curl/issues/17030}{curl/curl \#17030}.\par}

\end{SWETimeCaseCard}

\begin{SWETimeCaseCard}{R2}{cpython-86867}{Reject}

\emph{Insufficient symptom after sanitization.}

\textbf{Input --- regression report.} The issue is titled \emph{Discrepency between configure.ac and configure}. The extracted packet contains migration metadata and bisect information; the concrete \path{./configure} failure appears in a comment absent from the packet.

\textbf{Historical evidence.} The candidate commit listed below is identifiable, but its attribution overlaps with the limited technical content in the packet.

\textbf{Validation and decision.} Removing bisect and commit details leaves an underspecified query; retaining them exposes Gold. \textbf{Reject:} the sanitized packet lacks an independently intelligible regression symptom.

\par\smallskip{\footnotesize Candidate SHA: \path{c6d7e82d19c091af698d4e4b3623648e259843e3}.\par

Candidate ID: \path{python_cpython__86867__1bc451ca}.\par

Source: \href{https://github.com/python/cpython/issues/86867}{python/cpython \#86867}.\par}

\end{SWETimeCaseCard}

Together, these cases illustrate the admission boundary: a traceable developer first-bad attribution, a consistent Git revision, a behaviorally relevant transition, and a usable regression report are required for a Backward evaluation instance. Candidates failing attribution consistency or query sufficiency are excluded rather than assigned speculative Gold labels.

%% file: sections/english/appendix_b.tex
\setcounter{table}{0}
\setcounter{equation}{0}
\renewcommand{\thetable}{B.\arabic{table}}
\renewcommand{\theequation}{B.\arabic{equation}}
\renewcommand{\theHsection}{appendix.\Alph{section}}
\renewcommand{\theHsubsection}{appendix.\Alph{section}.\arabic{subsection}}
\renewcommand{\theHtable}{appendix.B.\arabic{table}}
\renewcommand{\theHequation}{appendix.B.\arabic{equation}}

\section{Forward Gold Construction and Validation}
\label{sec:appendix_foresight_gold}

Task Forward predicts the workflows activated by an intended repository change, given a real developer issue and the repository before its implementation. The historical fixing pull request (PR) supplies a reproducible Gold, but a request may admit multiple reasonable implementations. Deterministic routing and unambiguous task construction are therefore separate requirements. We describe the routing checks, developer review protocol, five worked examples, and independent audits used to assess this distinction.

\subsection{Gold Construction and Routing Validation}

Each candidate originates from a real GitHub issue linked to a merged fixing PR. Within the commit-indexed repository history $\mathcal{H}$ defined in Section~\ref{en:sec:benchmark_overview}, we recover the pre-change repository tree $S_t$ at the fixing PR's base commit and extract the paths $\Delta^*$ changed by that PR. We omit the case index $i$ in this appendix. Under a fixed routing context $E$, the activated-workflow Gold is
\begin{equation}
R^*=f(S_t,\Delta^*,E),
\label{eq:appendix_foresight_gold}
\end{equation}
where $R^*$ denotes the Gold workflow set, $S_t$ the pre-change repository state, $\Delta^*$ the historical changed-path set, $E$ the fixed routing context, and $f$ the frozen routing evaluator. Workflow activation follows Equation~\ref{en:eq:workflow_activation}. The benchmark fixes a non-draft pull-request synchronization context, including eligible \texttt{pull\_request} and \texttt{pull\_request\_target} triggers, direct activation, and the historical target branch. The evaluator uses workflow definitions in $S_t$, excluding workflow files introduced by the subsequent fix. Gold construction uses neither observed post-change CI executions nor an LLM judge. The agent receives only the sanitized issue $I$, pre-change state $S_t$, and context $E$, and predicts a workflow set through $h_\theta(I,S_t,E)$ as defined in Section~\ref{en:sec:task_forward}. The historical changed paths, fixing PR, and Gold are hidden from the agent; an explicit file-path prediction is not required.

The evaluator implements the activation predicates in Equation~\ref{en:eq:workflow_activation} from the workflow configurations in $S_t$. For each eligible event entry, $\phi_{\mathrm{event}}$ checks the configured event and activity type against $E$, including whether \texttt{types} admits \texttt{synchronize}. The branch predicate $\phi_{\mathrm{branch}}$ matches the target branch against \texttt{branches} or \texttt{branches-ignore}. The path predicate $\phi_{\mathrm{path}}$ accepts a \texttt{paths} filter if at least one changed path remains included, and a \texttt{paths-ignore} filter if at least one changed path is not ignored. Absent branch or path filters impose no restriction. Ordered include/exclude patterns preserve their precedence: a later matching negative pattern excludes a branch or path, while a later positive match re-includes it~\citep{github2026workflows}. An event entry activates the workflow only when its event, branch, and path predicates all hold; a workflow is selected if an eligible entry passes. The evaluator also resolves reusable-workflow relationships. The primary evaluation target remains the activated workflow set, rather than the historical PR's observed execution trace.

The frozen evaluator, \path{gold_routing_v4_1.effective_workflows}, is checked by a suite with independently hand-specified expected outputs. All 29 workflow-routing tests pass, covering event-action compatibility, target-branch filtering, and path-pattern behavior. This suite independently checks the specified routing semantics. Recomputing each stored Gold from its recorded $S_t,\Delta^*,E$ reproduces the accepted-case manifest and establishes reproducibility of the recorded labels.

\subsection{Human Review and Admission Protocol}

The admission criterion is \emph{workflow-level answerability}, rather than uniqueness of a fixing patch. For two developer-validated change scopes under the same historical context, we define routing equivalence as
\begin{equation}
\Delta_a\sim_{S_t,E}\Delta_b\iff R_a=R_b,
\label{eq:appendix_routing_equivalence}
\end{equation}
where $\Delta_a$ and $\Delta_b$ denote changed-path sets for two reasonable implementations, $S_t$ the fixed pre-change repository state, and $E$ the routing context. The workflow sets $R_a=f(S_t,\Delta_a,E)$ and $R_b=f(S_t,\Delta_b,E)$ are computed by the same frozen evaluator $f$; $\sim_{S_t,E}$ means that both scopes produce the same workflow set.

The review process assesses whether the issue and pre-change context sufficiently constrain the requested change, whether identified reasonable alternatives yield divergent workflow sets, and whether the historical PR and agent-visible inputs are consistent and free of answer leakage. A candidate can be admitted with multiple implementation locations when the reviewed alternatives remain routing-equivalent. A reasonable alternative yielding a different workflow set, insufficient evidence, or unresolved uncertainty precludes admission. Box~B.1 gives the operational rubric.

\begin{figure}[!t]
\centering
\setlength{\fboxsep}{6pt}
\fbox{\begin{minipage}{\dimexpr\columnwidth-2\fboxsep-2\fboxrule\relax}
\small
\textbf{Box B.1. Forward candidate review instructions.}\par\medskip
\textbf{Input.} A real developer issue $I$, pre-change repository tree $S_t$, frozen routing context $E$, and historical PR evidence for provenance and consistency checks. Inspect the agent-visible view separately from evidence reserved for Gold construction.

\begin{enumerate}
\setlength{\itemsep}{4pt}
\setlength{\parsep}{0pt}
\item \textbf{Scope sufficiency.} Does the issue, together with $S_t$, sufficiently constrain a credible modification direction? Identify the implicated components and supporting evidence. Do not demand a unique patch; assign Reject or Uncertain when the request is too underspecified to establish a defensible target.
\item \textbf{Workflow-level ambiguity.} Propose or inspect reasonable alternative implementations. For each credible changed-path set, evaluate the workflows under the same $S_t$ and $E$. Accept code-level variation when the workflow set is unchanged. Reject when a reasonable alternative yields a different target. Record the paths, rationale, and differing outputs; arbitrary paths are not alternative implementations.
\item \textbf{Historical consistency and leakage.} Verify that the fixing PR addresses the issue and that historical evidence supports the case. Check that the agent-visible input does not disclose the future implementation, changed paths, Gold, or other answer-bearing future evidence.
\end{enumerate}
\textbf{Decision.} Assign Accept, Reject, or Uncertain with supporting evidence. Do not admit unresolved cases. Refer disputed decisions to a third human adjudicator.
\end{minipage}}
\end{figure}

The screening considered 1,899 Forward candidates: 1,030 were accepted, 740 rejected, and 129 excluded as uncertain. Two senior software engineers independently annotated a separate reliability sample of 200 candidates using the three decision labels. Pre-adjudication agreement was 91.0\%, with Cohen's $\kappa=0.856$~\citep{cohen1960agreement}; a third reviewer adjudicated the 18 disagreements. Agreement is calculated from the reviewers' original labels, before adjudication.

\subsection{Worked Admission Cases: Inputs, Implementations, and Outputs}
\label{sec:appendix_admission_cases}

The following five examples make the admission rule concrete. Each specifies the developer-facing issue, relevant excerpts of historical routing rules, changed-path choices, activated workflows, and admission decision. The full configurations remain in the repository snapshots. Historical PRs establish actual changes; the alternative scopes shown here were separately reviewed by developers. A diagnostic routing probe tests a rule but need not implement the issue. For rejected examples, outputs are counterfactual evaluations of historical and alternative scopes; these candidates are excluded from the released benchmark.

\subsubsection*{Case A --- Constrained Modification Scope}

\begin{SWETimeCaseCard}{A1}{dyninst-2203}{Accept}
\textbf{Developer request.} In \path{dyninst/dyninst}, the issue names \texttt{Failed to transform trace <N>} near line 44 of \path{dyninstAPI/src/Relocation/Transformers/Transformer.C} and requests a more informative error message.

\textbf{Routing evidence.} Historical \path{pr-tests.yaml} admits PR actions \path{opened}, \path{synchronize}, \path{reopened}, and \path{ready_for_review} on \path{master}, with paths \path{**.h}, \path{**.C}, \path{**.c}, \path{**.cmake}, \path{**CMakeLists.txt}, and \path{tests/*}. The other three activated workflows use similarly broad C/C++ patterns.

\begin{SWETimeCaseOutputs}
Named \path{Transformer.C} statement (historical) & 4 workflows \\
Same fix plus a nearby header & Same 4 \\
\end{SWETimeCaseOutputs}
The workflow set is \{\path{binary-tests}, \path{codegen-cross-build}, \path{libabigail}, \path{pr-tests}\}. An unrelated \path{.C} file matches the routing filters but does not implement this issue; it is a diagnostic probe.

\textbf{Admission rationale.} The report directly identifies the emitting statement, and the reviewed fixes preserve the reproducible workflow Gold.
\SWETimeCaseSource{2204}{dyninst_dyninst__2203__af3e532}
\end{SWETimeCaseCard}

\begin{SWETimeCaseCard}{A2}{psutil-2639}{Accept}
\textbf{Developer request.} In \path{giampaolo/psutil}, an AIX source-distribution build of psutil v7.1.0 fails because \path{psutil/_psutil_posix.h} is missing. The packaging fix must make this header available.

\textbf{Routing evidence.} \path{build.yml} admits \path{push} and \path{pull_request} without a path filter; \path{bsd.yml} matches POSIX source paths; \path{issues.yml} admits the fixed PR event.

\begin{SWETimeCaseOutputs}
\path{psutil/_psutil_posix.h} (historical) & 3 workflows \\
Same header change plus \path{MANIFEST.in} & Same 3 \\
Same header change plus \path{setup.py} & Same 3 \\
\end{SWETimeCaseOutputs}
All reviewed scopes activate \{\path{bsd}, \path{build}, \path{issues}\}.

\textbf{Admission rationale.} The issue identifies a specific missing artifact and packaging objective. The historical fix and reviewed packaging variants are routing-equivalent.
\SWETimeCaseSource{2640}{giampaolo_psutil__2639__fb0792e}
\end{SWETimeCaseCard}

\subsubsection*{Case B --- Equivalent Workflow Gold}

\begin{SWETimeCaseCard}{B1}{fallow-2400}{Accept}
\textbf{Developer request.} ``Support SvelteKit v3'' in \path{fallow-rs/fallow} admits core-plugin, extraction-framework, or configuration implementations.

\textbf{Routing evidence.} On PRs to \path{main}, \path{fuzz-smoke.yml} admits \path{crates/extract/**}, \path{crates/core/**}, \path{crates/config/**}, and \path{crates/types/**}, among other paths.

\textbf{Scopes.} Paths below are relative to \path{crates/}.
\begin{SWETimeCaseOutputs}
\path{core/src/plugins/sveltekit.rs} plus integration tests/fixtures (historical) & 11 workflows \\
\path{extract/src/frameworks/sveltekit.rs} plus a core test & Same 11 \\
\path{config/src/frameworks.rs} plus the core plugin & Same 11 \\
\end{SWETimeCaseOutputs}
The common set is \{\path{allocs}, \path{bench}, \path{bloat}, \path{ci}, \path{commitlint}, \path{coupling}, \path{coverage}, \path{cross-arch}, \path{dependabot-auto-merge}, \path{ecosystem}, \path{fuzz-smoke}\}. A probe restricted to \path{crates/cli} and \path{crates/output} activates ten workflows, excluding \path{fuzz-smoke}; it is not a credible standalone implementation of the framework analyzer.

\textbf{Admission rationale.} The reviewed crate locations lie within the allowlist and preserve all 11 workflows; the accepted-case manifest and frozen evaluator agree.
\SWETimeCaseSource{2488}{fallow_rs_fallow__2400__75ca70b}
\end{SWETimeCaseCard}

\subsubsection*{Case C --- Divergent Workflow Gold}

\begin{SWETimeCaseCard}{C1}{data-prep-kit-1047}{Reject}
\textbf{Developer request.} ``HAP transform crashes when using a GPU'' can plausibly require a transform-local fix, shared processing-library fix, or GPU toolchain change.

\textbf{Routing evidence.} \path{test-universal-hap.yml} admits PRs to \path{dev} or \path{releases/**}, with \path{.make.*}, \path{transforms/universal/hap/**}, \path{data-processing-lib/**}, and \path{!**.md}. \path{test-processing-lib.yml} also matches \path{.make.*} and \path{data-processing-lib/**}, without a transform-local requirement.

\begin{SWETimeCaseOutputs}
\path{transforms/universal/hap/dpk_hap/transform.py} (historical) & 3 workflows \\
\path{data-processing-lib/**} & 33 workflows \\
\path{.make.versions} and \path{.make.defaults} & 57 workflows \\
\end{SWETimeCaseOutputs}
The historical set is \{\path{test-misc}, \path{test-universal-hap}, \path{test-universal-hap-kfp}\}. The shared scope adds \path{test-processing-lib} and sibling tests; toolchain changes activate further KFP and packaging workflows.

\textbf{Admission rationale.} The same request admits reasonable scopes with different workflow sets (3, 33, and 57). The candidate is excluded from the released benchmark.
\SWETimeCaseSource{1056}{data_prep_kit_data_prep_kit__1047__3245825}
\end{SWETimeCaseCard}

\begin{SWETimeCaseCard}{C2}{data-prep-kit-1015}{Reject}
\textbf{Developer request.} ``Improve performance of the Readability transform'' follows a slowdown on a large HuggingFace parquet workload. A fix could optimize the local transform or the shared \path{abstract_transform} base class.

\textbf{Routing evidence.} \path{test-language-readability.yml} matches \path{transforms/language/readability/**} and \path{data-processing-lib/**}. \path{test-processing-lib.yml} and roughly 30 related \path{test-*} workflows also match the shared-library paths.

\begin{SWETimeCaseOutputs}
\path{transforms/language/readability/**} (historical) & 2 workflows \\
\path{data-processing-lib/.../abstract_transform.py} & 33 workflows \\
\end{SWETimeCaseOutputs}
The historical set is \{\path{test-language-readability}, \path{test-misc}\}. The shared-library fix additionally activates \path{test-processing-lib} and other shared-library-dependent tests.

\textbf{Admission rationale.} The request does not determine local versus shared optimization, and the reasonable scopes yield different outputs (2 versus 33). The candidate is excluded from the released benchmark.
\SWETimeCaseSource{1026}{data_prep_kit_data_prep_kit__1015__54161c7}
\end{SWETimeCaseCard}

\subsection{Independent Audit of Accepted Cases}

We independently examined 100 previously accepted cases, sampling 25 from each Direct--Local, Direct--Cross, Staged--Local, and Staged--Cross stratum. This audit was separate from the original admission decisions. Reviewers reassessed scope sufficiency and whether reasonable alternative implementations changed the Gold.

The audit found 97 cases with no identified Gold divergence, one divergent case, and two inconclusive cases. Among the 98 conclusive decisions, 97 showed no identified divergence (99.0\%). Table~\ref{tab:appendix_foresight_audit} reports the stratum-level results. The divergent case and both inconclusive cases were removed and replaced by three newly admitted cases satisfying the full review criteria. The released Forward benchmark therefore retains 1,030 cases, while the original pre-replacement audit outcomes remain reported.

\begin{table}[!ht]
\centering
\small
\setlength{\tabcolsep}{3pt}
\resizebox{\columnwidth}{!}{%
\begin{tabular}{@{}lrrrr@{}}
\toprule
\textbf{Subset} & \textbf{Audited} & \makecell{\textbf{No}\\\textbf{divergence}} & \textbf{Divergent} & \textbf{Inconclusive} \\
\midrule
Direct--Local & 25 & 25 & 0 & 0 \\
Direct--Cross & 25 & 24 & 0 & 1 \\
Staged--Local & 25 & 24 & 1 & 0 \\
Staged--Cross & 25 & 24 & 0 & 1 \\
\midrule
\textbf{Total} & \textbf{100} & \textbf{97} & \textbf{1} & \textbf{2} \\
\bottomrule
\end{tabular}
}
\caption{Independent audit of previously accepted Task Forward cases.}
\label{tab:appendix_foresight_audit}
\end{table}

\subsection{Alternative Implementation Audit and Scoring Sensitivity}

In the same 100-case audit, developers proposed and reviewed alternative implementation scopes grounded in each issue and its historical repository. 34 cases had at least one developer-validated alternative scope suitable for routing comparison. For 33 cases, the alternative preserved the historical Gold; one produced a different workflow set, yielding 97.1\% alternative-scope Gold agreement. The worked cases in Section~\ref{sec:appendix_admission_cases} illustrate stable and divergent routing. Its two rejected Case C examples are construction-stage candidates, separate from the single divergent case in this independent audit.

We also measure scoring sensitivity to reasonable alternative targets. On the original 100-case pre-replacement audit sample, the reference agent obtains 63.0\% Overall against historical Gold and 64.0\% Overall when validated alternative Gold sets are additionally accepted, a gain of 1.0 percentage point. This analysis holds predictions fixed and varies the accepted Gold targets. The balanced audit describes the examined sample and its validated alternatives; the benchmark's primary historical-Gold scoring protocol remains unchanged.

Together, routing validation establishes reproducible labels, human review excludes identified workflow ambiguities, and independent audits detect missed cases and guide corrections before release.

%% file: sections/english/appendix_c.tex
\setcounter{table}{0}
\setcounter{equation}{0}
\renewcommand{\thetable}{C.\arabic{table}}
\renewcommand{\theequation}{C.\arabic{equation}}
\renewcommand{\theHtable}{appendix.C.\arabic{table}}
\renewcommand{\theHequation}{appendix.C.\arabic{equation}}
\section{Historical Information Isolation and Reproducibility}
\label{sec:appendix_historical_isolation}

SWE-Time constructs verifiable targets from later repository evidence while evaluating agents at an earlier information boundary. Construction and agent execution use separate environments. This appendix describes historical views, leakage controls, evaluation checks, diagnostic labels, and task instructions; Appendices~\ref{sec:appendix_forensics_gold} and~\ref{sec:appendix_foresight_gold} detail Gold attribution and routing validation, respectively.

\subsection{Task-Specific Historical Views}
\label{sec:appendix_historical_views}

\paragraph{Backward: cutoff-restricted Git history.}
Each instance contains a sanitized regression report and Git history ending at a case-specific cutoff derived from the report's historical context. The pipeline resolves the developer-attributed first-bad SHA, checks ancestry and first-parent branch consistency, and materializes an isolated repository retaining original commit identities and permitted historical states. Under Full History, agents may inspect permitted revisions and diffs, build or test code, and create diagnostic probes. Developer bisect attributions and construction-time good/bad probes remain hidden.

\paragraph{Historical metadata and evidence removal.}
The pre-cutoff candidate history includes \path{history_metadata.jsonl}, recording each commit's SHA, parents, timestamp, author, and message. Both Full History and the Metadata-Only condition in RQ2 expose these fields. Full History additionally exposes historical tree and blob objects for code inspection and diff computation; Metadata-Only removes historical code states and diffs while preserving commit sequence and metadata. In RQ1, BM25 (commit message) uses pre-cutoff messages, whereas BM25 (diff / paths) uses changed paths and patch content, which the Metadata-Only agent cannot access.

Reports are sanitized by \emph{deletion rather than paraphrasing}: culprit SHAs, bisect results, commit or pull-request links, and equivalent answer-bearing spans are removed; reproduction steps, error messages, and substantive symptoms remain. Auxiliary repository materials, including release notes and documents identifying the culprit, are also screened in the agent-visible view. Historical code and changes remain available for investigation.

\paragraph{Forward: pre-change working tree.}
Each instance exposes the developer issue, the repository state $S_t$ reconstructed from the fixing PR's base commit, and a fixed evaluation context $E$. The tree is extracted from the historical repository mirror without a \texttt{.git} directory. The actual fixing PR, its patch, the changed-path set $\Delta^*$, and subsequent workflow definitions remain construction-only evidence. The canonical context represents a non-draft pull-request \texttt{synchronize} event, with the target branch resolved at $S_t$ and provided consistently to both the agent and the routing evaluator. Gold is computed from workflow definitions already present in $S_t$.

\subsection{Packaging and Leakage Checks}
\label{sec:appendix_packaging_checks}

Each Backward package has a dedicated evaluation ref. Before admission, the audit checks cutoff availability, Gold membership in the permitted history, inaccessibility of the immediate post-cutoff first-parent probe, and visible-tree consistency with its recorded digest. It also checks for unintended refs and shallow-history artifacts. Packages from the same repository are materialized serially to avoid cross-instance contamination.

Forward packages archive the pre-change tree into the execution image without a Git object database. Snapshot integrity is checked against the historical tree used to compute Gold. The canonical changed-path list and routing target remain hidden for construction and grading. Paths naturally mentioned in the issue remain visible even when they overlap with the fix. Both tasks supplement automatic checks with case-level review for indirect answer leakage.

The runner uses network/DNS restrictions and proxy overrides to block GitHub and other external sources of future issue, commit, or PR evidence while permitting required evaluation infrastructure connections. Network-isolation checks are recorded with each trial.

\subsection{Agent Execution and Scoring}
\label{sec:appendix_agent_execution}

The benchmark is executed in per-case containerized tasks using Claude Code v2.1.220 and Codex v0.112.0. Both harnesses have a 3,600-second trial timeout; Codex uses \path{reasoning_effort="high"} and \path{override_timeout_sec=3600}. Automatic trial retries are disabled (\path{max_retries=0}); coverage backfills use separate invocations of the evaluation runner. Comparisons within the same harness and evaluation condition use common task instructions, tool-access policies, and inference-budget policies. Backward permits repository inspection, Git operations, local execution, and agent-authored tests or probes, while the construction-time regression oracle remains hidden. Expiration without a correct submission is unsuccessful. The Backward grader and answer-bearing task metadata are kept outside the agent workspace; grading artifacts are installed only after the agent run. Forward requires predictions of activated workflows, without executing prospective fixes or triggering actual CI runs.

Scoring is deterministic and independent of the agent's explanation. Backward accepts a complete 40-character commit SHA and compares it with the stored first-bad Gold. Forward compares predicted workflow-path sets against the frozen routing Gold. Missing or extra workflow identifiers prevent an Exact match. A well-formed empty workflow set is a scorable prediction and is evaluated by the same set comparison. The Forward evaluator resolves event/action eligibility, target-branch filters, and changed-path include/exclude patterns. The event context, evaluator semantics, case selection, and scoring rules are fixed independently of model predictions.

\subsection{Preflight Validation and Result Aggregation}
\label{sec:appendix_preflight_aggregation}

Before model evaluation, task packages and scorers undergo complementary checks: an \emph{oracle} submission containing the stored Gold must receive full credit, while malformed submissions and valid submissions that differ from Gold must receive no credit. These tests validate agreement among the packaged task, hidden target, output parser, and scorer. Independent historical and behavioral Gold validation is described in Appendix~\ref{sec:appendix_forensics_gold}; hand-specified workflow-routing tests validate evaluator semantics in Appendix~\ref{sec:appendix_foresight_gold}.

Each trial is associated with its canonical case identifier. Task and protocol manifests associate the input state, frozen evaluation context, and Gold with the corresponding result. Reported Overall scores are case-weighted Exact over the relevant task set; timeouts, missing submissions, and unscorable outputs receive zero and remain in the denominator. The SWE-Time average gives Backward and Forward equal weight. Controlled experiments reuse fixed diagnostic subsets, whereas full-task analyses use the complete admitted sets.

In paired-state analyses, semantically invalid candidate pairs are excluded before evaluation. Pair Exact requires correct predictions in both states and is computed over all retained pairs, with any unscorable output counted as a failure. Invariance Error is the proportion of pairs with unchanged predictions despite different Gold targets, computed only among pairs with scorable outputs in both states. Invalid is the percentage of retained pairs containing at least one unscorable output. Gold-construction artifacts, agent-visible inputs, and hidden scoring data are stored and accessed separately.

\subsection{Diagnostic Label Definitions}
\label{sec:appendix_diagnostic_labels}

Temporal Low and Temporal High correspond to Direct and Staged, respectively. These labels are assigned independently of model performance and frozen before evaluation. Their task-specific criteria describe the reasoning required to recover the target.

\paragraph{Backward.}
Local/Cross compares the component implicated by the regression symptom with the component implicated by the Gold change: Local denotes the same component, and Cross denotes different components. Direct denotes a clear historical boundary; Staged denotes cases with credible earlier fix-like changes that must be distinguished when localizing the regression origin. This distinction concerns historical localization difficulty; Gold admission separately requires a resolved developer attribution and behavioral validation.

\paragraph{Forward.}
Local denotes a historical changed-path set confined to one top-level repository component; Cross spans at least two such components. Temporal labels are assigned through offline inspection of the pre-change snapshot. Direct corresponds to \emph{explicit routing}, where the request and repository evidence support direct application of workflow rules. Staged covers \emph{scope grounding} or \emph{impact escalation}, requiring further inference of the modification scope or its effects before workflow activation can be determined.

\subsection{Task Instruction Examples}
\label{sec:appendix_task_instructions}

The following excerpts show the task inputs, evaluation context, and answer format. Long report logs are abbreviated, and the Forward example presents the workflow-level instructions and response fields.

\begin{SWETimeCaseCard}{Prompt C.1}{Backward}{Excerpt}
\textbf{Task.} Locate the commit that introduced this regression.

\textbf{Report.} Assignment to uint series drops datatype. Assigning a uint value to a slice of a uint series results in a dataframe of type int. With Pandas 0.19.2 the series retains \path{uint64}; with version 0.21.0 it becomes \path{int64}.

\textbf{What you have.} \path{/testbed/repo} is a Git clone of \path{pandas-dev/pandas} truncated just after the regression was reported. The commit that introduced the regression is an ancestor of HEAD; the later fixing commit is absent. Reading history, \path{git log}, \path{git diff}, \path{git show}, and writing probe scripts support investigation.

\textbf{What to produce.} Write the full SHA of the first-bad commit into \path{/testbed/answer.txt}: exactly one line of 40 lowercase hexadecimal characters, with no explanation or surrounding punctuation. Abbreviated SHAs are rejected, and the commit must exist in the provided repository.

\textbf{Rules.} Do not use the network, rewrite history, add remotes, or fetch objects. Solve the task from the provided repository.
\end{SWETimeCaseCard}

\begin{SWETimeCaseCard}{Prompt C.2}{Forward}{Excerpt}
\textbf{Task.} Recover the symbolic verification state this change produces.

\textbf{Report.} In \path{giampaolo/psutil}, an AIX source-distribution build of psutil v7.1.0 fails because \path{psutil/_psutil_posix.h} is missing.

\textbf{What you have.} \path{/testbed/repo} contains the pre-change working tree, without Git history. The PR that actually fixed the issue is absent and cannot be fetched.

\textbf{Evaluation context.} Assume the prospective fix has updated an existing non-draft PR: action \path{synchronize}, target branch \path{master}. Consider \path{pull_request} and \path{pull_request_target}; apply native \path{types}, \path{branches}, \path{branches-ignore}, \path{paths}, and \path{paths-ignore}. Consider direct workflow activation, excluding \path{push}, \path{workflow_dispatch}, \path{schedule}, and \path{workflow_run}.

\textbf{What to produce.} Write JSON to \path{/testbed/answer.json}, with the case identifier and predicted workflow paths:

{\ttfamily\footnotesize
\{\textnormal{\texttt{"case\_id"}}: "<case-id>",\par
\hspace*{1em}"activated\_workflows": [\par
\hspace*{2em}"<workflow-path>"\par
\hspace*{1em}]\}\par}
Workflow paths must use their exact locations under \path{.github/workflows/} in the provided snapshot.

\textbf{Rules.} Do not use the network or fetch the upstream fix. Solve the task from the provided repository.
\end{SWETimeCaseCard}

\subsection{Controlled Input Interventions}
\label{sec:appendix_controlled_inputs}

RQ1 baselines and RQ2--RQ3 use a fixed 300-case diagnostic subset for each task. Controlled conditions within a task reuse the same cases. Full Report and Full Issue retain the base instructions; the query interventions below edit the instruction text while reusing the corresponding repository image and task specification.

\paragraph{Symptom-Ablated.}
For Backward, symptom-defining spans and title phrasing are deleted from the sanitized report. The remaining report content is retained verbatim, and the available historical search space is unchanged.

\paragraph{No Issue.}
For Forward, the issue-introduction line and the complete report block are removed. Repository access, routing context, task instructions, and the response format remain the same.

\paragraph{Oracle Scope.}
For Forward, an \emph{Actual change scope} block containing the normalized, sorted historical changed-path set $\Delta^*$ is inserted before the repository-access instructions. The issue, pre-change snapshot, and routing context are unchanged. This diagnostic condition supplies change-scope information in addition to the base inputs.

The Metadata-Only intervention removes historical code evidence as described in Section~\ref{sec:appendix_historical_views}. Paired-state contrasts retain the issue, historical change scope, and routing context while replacing the repository state, as described in RQ2.

%% file: sections/english/appendix_d.tex
\setcounter{table}{0}
\setcounter{equation}{0}
\renewcommand{\thetable}{D.\arabic{table}}
\renewcommand{\theequation}{D.\arabic{equation}}
\renewcommand{\theHtable}{appendix.D.\arabic{table}}
\renewcommand{\theHequation}{appendix.D.\arabic{equation}}
\section{Repository-Level Robustness}
\label{sec:appendix_repository_robustness}

We assess sensitivity to repository weighting and single-repository removal using full-task predictions. Table~\ref{tab:appendix_repository_robustness} reports Task Forward results for four models under Claude Code and GPT 5.4 under Codex as a reference; model comparisons remain within the Claude Code group. Case-weighted Overall averages case-level correctness, whereas Repo-Macro first computes performance within each repository and then weights repositories equally. Leave-one-repository-out (LOO) evaluation removes each of the 16 repositories in turn and recomputes case-weighted Overall on the remaining cases.

\begin{table}[!ht]
\centering
\begingroup
\small
\setlength{\tabcolsep}{2pt}
\renewcommand{\arraystretch}{1.08}
\begin{tabular*}{\columnwidth}{@{\extracolsep{\fill}}lcccc@{}}
\toprule
\textbf{Model} & \makecell{\textbf{Case-}\\\textbf{weighted}} & \makecell{\textbf{Repo-}\\\textbf{Macro}} & \makecell{\textbf{LOO}\\\textbf{Range}} & \makecell{\textbf{Max.}\\\textbf{Shift}} \\
\midrule
Qwen 3.8 Max & 63.9 & 61.5 & 59.5--71.2 & 7.3 \\
Claude Opus 4.8 & 62.7 & 64.6 & 58.1--69.9 & 7.2 \\
GLM 5.2 & 58.3 & 58.1 & 53.8--65.0 & 6.7 \\
DeepSeek V4 Flash & 46.2 & 47.8 & 41.5--51.5 & 5.3 \\
\midrule
GPT 5.4 & 61.7 & 61.6 & 56.8--68.7 & 7.1 \\
\bottomrule
\end{tabular*}
\endgroup
\caption{\textbf{Repository-level robustness on Task Forward.} Scores and LOO ranges report Overall (\%). LOO Range gives the minimum and maximum scores after single-repository removal; Max. Shift gives the largest absolute change from the full-task score in percentage points. GPT 5.4 is a Codex reference; the other models use Claude Code.}
\label{tab:appendix_repository_robustness}
\end{table}

The small differences between case-weighted and repository-macro scores indicate that the substantial performance headroom persists under equal repository weighting. Across the four Claude Code models, unchanged ordering in all 16 LOO evaluations further shows that their relative performance is robust to removing any single repository. Absolute scores nevertheless remain sensitive to repository composition, and the Opus--Qwen ordering reverses under repository-macro aggregation.

\paragraph{Repository-composition uncertainty.}
For Claude Opus 4.8, repository-clustered bootstrap~\citep{field2007clustered,efron1979bootstrap} gives 95\% confidence intervals (CIs) for case-weighted Overall of [55.4, 66.9] for Backward and [49.4, 71.6] for Forward. We use 10,000 resamples and random seed 20261011. Within each task, we draw repositories uniformly with replacement, drawing the original number of repositories (49 for Backward and 16 for Forward). Each selected repository contributes all its case outcomes, with repeated contributions when sampled more than once. We recompute case-weighted Overall from the pooled correct and total counts and take the 2.5th and 97.5th percentiles as the interval endpoints~\citep{diciccio1996bootstrap}. These intervals quantify uncertainty under repository-level resampling; the wider Forward interval indicates greater sensitivity to repository composition.